\documentstyle[12pt,aaspp4]{article}
\def\lsim{\lower.5ex\hbox{$\; \buildrel < \over \sim \;$}}
\def\gsim{\lower.5ex\hbox{$\; \buildrel > \over \sim \;$}}
\lefthead{}
\righthead{Unification to the analysis of accretion disks around rotating compact objects}

\def\lsim{\lower.5ex\hbox{$\; \buildrel < \over \sim \;$}}
\def\gsim{\lower.5ex\hbox{$\; \buildrel > \over \sim \;$}}

\received{2002 September 8}
\begin{document}

\title{Unification to the {\bf\it Pseudo-General-Relativistic} Analysis of Accretion Disks around Rotating 
Black Holes and Neutron Stars}
 
\author{Banibrata Mukhopadhyay}

\affil{\small Inter-University Centre for Astronomy and Astrophysics,
Post Bag 4, Ganeshkhind, Pune-411007, India\\
}

\begin{abstract}

I analyse the relativistic accretion phenomena around rotating black holes and neutron stars
and show both the kinds of disk can be treated in an unified manner.
The corresponding accretion disks are described by pseudo-Newtonian approach.
For this purpose, number of pseudo-Newtonian potentials are in literature,
applicable to describe the relativistic properties of accretion disk. While, Kerr metric is used
to describe the pseudo-Newtonian potential for accretion disk around black hole,
the Hartle-Thorne metric is considered to describe disk around neutron star as the metric
can describe continuously the space-time, inside the star as well as out-side of it. 
Two other potentials were proposed to describe 
the temporal effects of the accretion disk. All the potentials reproduce
the marginally stable and bound orbits approximately or exactly as that of general relativity. 
These also reproduce the specific mechanical energy approximately. Using these potentials, I study the global
parameter space of the accretion disk around black holes and neutron stars. 
I study, how the fluid properties get affected for different angular momentum of the compact object.
I show that, for different angular momenta of the compact object, the valid disk parameter region dramatically 
changes and disk may become unstable in certain situations. Also I discuss about the possibility of 
shock in accretion disk around rotating black holes and neutron stars. When the angular momentum of compact object is chosen 
to be varied, the sonic locations of the accretion disk get shifted or disappear, making the disk unstable.
To bring it in the stable situation, the angular momentum
of the accreting matter has to be reduced/enhanced (for co/counter-rotating disk)
by means of some physical process. I also study, how the fluid properties get changed with 
different rotations of the black holes, neutron stars and other gravitating central stars. 
Moreover, I show the effect of viscosity to the
fluid properties of the disk. Thus, I find out the unified physical parameter regime, for which the stable 
accretion disk can be formed. Subsequently, a theoretical prediction of kHz QPO is given,
for a fast rotating compact object as 4U 1636-53.

\end{abstract}

\keywords {accretion, accretion disk --- black hole physics --- stars: neutron --- 
hydrodynamics --- shock waves --- gravitation }

\section{Introduction}

One of the important subject in astrophysics is the study of fluid dynamical phenomena 
of the accretion disk around compact object. There are three main kinds of compact objects, namely, white dwarf,
neutron star and black hole. Also in last decade, the possibility of the strange compact star contained by the 
quark matter, has been proposed. If the compact object is a companion of a binary system or situated at the
centre of any galaxy (if the compact object is black hole), the accretion disk forms around it.
Shakura \& Sunyaev (1973) started this discussion theoretically,
proposing a model of accretion disk, namely Keplerian accretion disk model. Further,
Novikov \& Thorne (1973) and Page \& Thorne (1974) analysed accretion disk in a
general relativistic treatment. Paczy\'nski \& Wiita (1980), for first time, proposed a method namely 
{\it pseudo-Newtonian} and prescribed the corresponding pseudo-Newtonian potential to describe the accretion disk
around Schwarzschild (non-rotating) black holes. In this method, the basic equations of the disk are chosen 
as non-relativistic and the gravitational force and corresponding potential are adjusted in such a manner, 
that could approximately describe 
relativistic properties of the accretion disk around compact objects. Using that pseudo-Newtonian potential for
non-rotating compact, Abramowicz \& Zurek (1981) studied some transonic 
aspects of accretion phenomena. However, as no cosmic object is static, the pseudo-Newtonian potential
proposed by Paczy\'nski \& Wiita (1980, hereinafter PW potential) is oversimplified, particularly for
inner region of the accretion disk. Thus other pseudo-Newtonian potentials have been proposed incorporating the
rotation of the compact object by Artemova et al. (1996), Mukhopadhyay (2002a), Mukhopadhyay \& Misra
(2003), Mukhopadhyay \& Ghosh (2003) and Ghosh (2003). The interesting fact lies in the methodology adopted 
by Mukhopadhyay (2002a) to describe a potential for the accretion disk, which can be used to derive the 
pseudo-potential for any metric according to the physics concerned.

Out of several works on accretion phenomena present in literature, mostly are discussed for black holes 
(e.g., Abramowicz et al. 1988, 1996; Abramowicz \& Kato 1989; Chakrabarti 1990, 1996a,b,c; Narayan \& Yi 1994; 
Narayan et al. 1997, 1998; Kato et al. 1998; Sponholz \& Molteni 1994;
Peitz \& Appl 1997; Gammie \& Popham 1998; Popham \& Gammie 1998; Miwa et al. 1998;
Lu \& Yuan 1998; Manmoto 2000; Mukhopadhyay 2003) and a few are for neutron stars
(Chakrabarti \& Sahu 1997; Mukhopadhyay 2000, 2002b; Popham \& Sunyaev 2001; Prasanna \& Mukhopadhyay 2003;
Mukhopadhyay \& Ghosh 2003). These studies have been made either in a full general relativistic frame-work or in a 
pseudo-Newtonian approach. There are a number of basic differences between the accretion disk around black hole
and any other gravitating objects namely neutron star, white dwarf, strange star and other highly
gravitating star, particularly at the inner edge of the disk.
(1) While the black hole has event horizon, other gravitating objects mentioned above have
hard surface. Thus, whereas in the case of other gravitating object close to the surface
the matter speed must be subsonic, for black holes the matter speed is supersonic.
(2) In case of other gravitating object that has hard surface, the matter comes into contact with
its outer surface. Thus, there is an obvious effect of dragging to the exterior space-time
due to its rotation. In terms of general relativity, corresponding space-time metric
should be the solution for interior as well as exterior to it. Therefore, it is
necessary to consider the 'dragging effect' of rotating gravitating star (particularly for
neutron star, where the relativistic effect is very important) to the space-time.
In case of black hole, which does not have hard surface, this effect comes to the space-time
automatically through the 'Lense-Thirring drag' from the Kerr geometry which does not
describe any physics interior to the horizon.
(3) Further, the overall temperature as well as the luminosity of the
disk is expected to be high in the cases of other gravitating star particularly for
neutron star with respect to that of black hole (Mukhopadhyay 2000, 2002b; Mukhopadhyay \& Ghosh 2003) 
for a particular accretion rate. If the same parameters are chosen for the infalling matter towards
black hole and other gravitating objects separately, at a particular accretion rate, the maximum
temperature in the disk around black hole will be lower at least a few factors times.
Thus, in the disk around a neutron star, expected generation/absorption of nuclear energy
is higher than that around a black hole. As the radius of white dwarf
is very large, the general relativity is not much important in the accretion disk around it.
But my discussions will be generally true for disk around black hole and any other kind of
gravitating object having hard surface, namely neutron star, white dwarf,
strange star and other gravitating star.

Mukhopadhyay (2002b) has studied the fluid behaviour of accretion disk around
non-rotating neutron star with the negligible magnetic effect compared to the viscous effect in
the disk. It was found that the temperature of the inner disk around neutron star 
is significantly higher than the case of
black holes. Prasanna \& Mukhopadhyay (2003) have shown that the rotational effect of compact object 
(particularly other than black hole) to the accretion disk can be approximately reproduced with the use
of Coriolis effect. They have done perturbative analysis and shown that the
self-similar solutions are well behaved for most of the parameter space.
Moreover, the recent observation tells about the evidence of millisecond pulser
(Strohmayer \& Markwardt 2002) of frequency 582Hz for 4U 1636-53 which is the accreting 
Low-Mass-X-ray-Binary (in short LMXB).
Therefore, it is important to consider properly the rotational effect of the neutron star and 
strange star to the disk. Apart from that, it is established fact that the inner edge properties
of accretion disk around black hole strongly depend on the angular momentum of black hole
(e.g., Gammie \& Popham 1998; Popham \& Gammie 1998; Mukhopadhyay 2003).

In this paper, I shall study the accretion phenomena in pseudo-Newtonian manner mainly following potentials
given by Mukhopadhyay (2002, hereinafter Mukhopadhyay potential), Mukhopadhyay \& Misra (2003, hereinafter
MM potential) and Mukhopadhyay \& Ghosh (2003, hereinafter MG potential). 
As using those pseudo-potentials, I shall study the relativistic properties of accretion disk (using
simple non-relativistic set of fluid equations), I name this study as {\it pseudo-general-relativistic}. 
Moreover, I will study the viscous accretion phenomena for entire physical parameter region and
comment on the stability of the disk. I also will analyse globally the behaviour of the sonic points,
formation of shock and show instead of several differences, the accretion disk around black holes
and neutron stars, both can be studied in an equal footing and all of their properties belong to
an unified class. Recently, Mukhopadhyay et al. (2003)
prescribed a theoretical model to achieve the QPO frequency for LMXB for rotating compact object and 
showed their theoretical prediction of frequency matches fairly well with that of observation. 
As my main interest is to study the inner region of the accretion disk, I shall concentrate upon 
the sub-Keplerian flow, where the effect of rotation of compact object is 
significant. 

In the next section, I will present the basic equations needed to describe the accretion disk. 
In \S 3, I will discuss about the parameter space of accretion disk and how does the rotation 
of compact object affect it.
Subsequently, in \S 4, I will describe the fluid dynamical results. Finally, in
\S 5, I will sum up all the results and make the overall conclusions.

\section{Basic model equations of accretion disk  }\label{sec:ref}

In the calculations, throughout I express the radial coordinate
in unit of $GM/c^2$, where $M$ is mass of the compact object, $G$ is the gravitational constant and
$c$ is the speed of light. I also express the velocity in unit of speed of
light and the specific angular momentum in unit of $GM/c$. The equations to be solved are given below.\\
\noindent (1) Equation of continuity:
   \begin{eqnarray}
   \nonumber
  \frac{1}{x} \frac{d}{dx}\left(x\Sigma v\right)=0\\
   {\rm or}\hskip0.2cm
  \dot{M}= -4\pi x\Sigma v,
   \label{ec}
   \end{eqnarray}
   where $\Sigma$ is the vertically integrated density and $\dot{M}$ is the accretion rate. 
Following Matsumoto et al. (1984), I can calculate $\Sigma$ as
\begin{equation}
  \Sigma=I_n\rho_e h(x), 
  \label{den}
  \end{equation}
  where $\rho_e$ is the density at equatorial plane, $h(x)$ is the half-thickness of the disk and
  $I_n=\frac{(2^n n!)^2}{(2n+1)!}$, where $n$ is the polytropic index. From the vertical equilibrium 
  assumption, the half-thickness can be written as
  \begin{equation}
   h(x)=c_s x^{1/2}F^{-1/2},
  \label{ht}
  \end{equation}
  where $c_s$ is the speed of sound. $F$ is the pseudo-Newtonian gravitational force. Its form will
  depend on the nature of compact object. If the compact object is non-rotating it will be according
  to PW potential. If the compact object is rotating black hole or neutron star, its expression will
  be in terms of Mukhopadhyay potential or MG potential respectively.\\ 
  \noindent (2) Radial momentum balance equation:
   \begin{equation}
   v\frac{dv}{dx}+\frac{1}{\rho}\frac{dP}{dx}-\frac{\lambda^2}{x^3}+F(x)=0, 
   \label{rmom}
   \end{equation}
   where the flow is chosen as adiabatic with the equation of state to be $P=K\rho^\gamma$ and $c_s^2=\frac{\gamma P}{\rho}$.
   If the compact object is non-rotating
   \begin{equation}
   F(x)=F_{\rm PW}= \frac{1}{(x-2)^2}. 
   \label{pw}
   \end{equation}
   For rotating black hole
   \begin{equation}
      F(x)=F_{\rm BH}= \frac{(x^2-2J\sqrt{x}+J^2)^2}{x^3(\sqrt{x}(x-2)+J)^2},
      \label{M}
   \end{equation}
and for rotating neutron star
    \begin{equation}
       F(x)=F_{\rm NS}=\frac{ x^3(x^3-4J^2)^2}{[J x^3(3x-4)-2J^3 +\sqrt{(5J^2+x^3)(2J^2+x^3(x-2))^2}]^2}.
      \label{MG}
   \end{equation}
Also following MM, for the time-dependent study of accretion phenomena 
\begin{equation}
 F(x)=F_{\rm MM}={1\over x^2}\left[ 1 - \left({x_{s}\over x}\right) + \left({x_{s}\over
 x}\right)^2\right],
       \label{MM} 
   \end{equation}
where $x_{s}$ is given by as
\begin{eqnarray}
x_{s} & = &3 + Z_2 \mp[(3-Z_1)(3+Z_1+2Z_2)]^{1/2} \\
\nonumber
Z_1 & = & 1 +(1-J^2)^{1/3}[(1+J)^{1/3}+(1-J)^{1/3}] \\
\nonumber
Z_2 & = & (3J^2+Z_1^2)^{1/2},
\end{eqnarray}
'-' ('+') sign is for the co-rotating (counter-rotating) flow.
The parameter, $J$, is the specific angular momentum of the compact object.\\
   \noindent (3) Azimuthal momentum balance equation:
   Here, I will follow Chakrabarti (1996a) to express the viscous dissipation $Q^+$ in terms of shear stress
   $W_{x\phi}$. If the viscous dissipation vanishes, the angular momentum of accreting matter remains constant
   throughout the disk. 
   Thus, $W_{x\phi}=-\alpha(I_{n+1}P+I_n v^2\rho)h(x)$ and $Q^+=\frac{W_{x\phi}^2}{\eta}$, $\eta$ is 
   coefficient of viscosity and $\alpha$ is Shakura-Sunyaev (1973) viscosity parameter. Thus the equation to be
   \begin{equation}
   v\frac{d\lambda}{dx}=\frac{1}{\rho h(x) x}\frac{d}{dx}\left[x^2\alpha\left(\frac{I_{n+1}}{I_n}P+v^2\rho\right)h(x)\right].
   \label{azmom}
   \end{equation}
(4) Entropy equation: According to the mixed shear stress (Chakrabarti 1996a), 
$Q^+=-\alpha(I_{n+1}P+I_n v^2\rho)h(x)x\frac{d\Omega}{dx}$. For simplicity, I also consider the heat lost
is proportional to the heat gained by the flow. Thus the equation to be
   \begin{equation}
   \Sigma vT\frac{ds}{dx}=\frac{vh(x)}{\Gamma_3-1}\left(\frac{dP}{dx}-\Gamma_1\frac{P}{\rho}\frac{d\rho}{dx}\right)=Q^+-Q^-=fQ^+,
   \label{enteq}
   \end{equation}
  where $s$ is entropy density and $f$ is cooling factor which is close to $0$ and $1$ for the flow with efficient 
  and inefficient cooling respectively. Following Cox \& Giuli (1968), I can define
  \begin{equation}
  \nonumber
   \Gamma_3=1+\frac{\Gamma_1-\beta}{4-3\beta},\\
   \nonumber
   \Gamma_1=\beta+\frac{(4-3\beta)^2(\gamma-1)}{\beta+12(\gamma-1)(1-\beta)},\\
   \beta=\frac{\rho kT/\mu m_p}{\bar{a} T^4/3+\rho k T/\mu m_p}.
   \label{gambet}
   \end{equation}
   Here, $\beta$ (ratio of gas pressure to total pressure) close to $0$ for radiation 
   dominated highly relativistic flow and close to $1$ for
   gas dominated flow.
Now combining (\ref{ec})-(\ref{enteq}), I get
\begin{equation}
\frac{dv}{dx}=\frac{f_1(x,v,c_s)}{f_2(v,c_s)},
\label{dvdx}
\end{equation}
where,
\begin{eqnarray}
\nonumber
f_1(x,v,c_s)&=&\left[c_s^2\left(\frac{3}{2x}-\frac{1}{2F}\frac{dF}{dx}\right)-\gamma\left(F-\frac{\lambda^2}{x^3}\right)\right]
\left[v(\Gamma_1+1)-\frac{4\alpha\Lambda}{v(3\gamma-1)}\right] \\
&-&\left[\frac{2\alpha c_s^2}{(3\gamma-1)xv}+
\frac{\alpha v}{x}-\frac{2\lambda}{x^2}\right]\Lambda+\left(\frac{3}{2x}-\frac{1}{2F}\frac{dF}{dx}\right)
v c_s^2(\Gamma_1-1),
\label{f1}
\end{eqnarray}
\begin{eqnarray}
f_2(v,c_s)=\left[1-\frac{2c_s^2}{(3\gamma-1)v^2}\right]\alpha\Lambda-c_s^2(\Gamma_1-1)
+\left(\gamma v-\frac{c_s^2}{v}\right)\left[v(\Gamma_1+1)-\frac{4\alpha\Lambda}{(3\gamma-1)v}\right]
\label{f2}
\end{eqnarray}
and $\Lambda=\gamma(\Gamma_3-1)f\alpha\left(\frac{I_{n+1}}{I_n}\frac{c_s^2}{\gamma}+v^2\right)$.\\
If the compact object is black hole, matter speed must be supersonic close to it. On the other hand,
for neutron star matter speed may or may not attain supersonic speed throughout its path in the 
accretion disk, but very close to the compact object, matter speed must be subsonic as star has hard surface (Mukhopadhyay 2002b).
Thus although at far away from the compact object, $v<c_s$, close to it, for black hole always
and for neutron stars in some cases, $v>c_s$. Therefore, in that situation, there must be an intermediate
location, where the denominator of (\ref{dvdx}) must be vanished. To have a smooth solution, at that
location, numerator has to be zero. This location is called the sonic point or critical point ($x_c$). The existence of 
sonic location plays an important role in accretion phenomena. 
Sonic point is the main factor behind the formation of shock in accretion disk.
From the global study of the sonic points in an accretion disk, one can 
understand about the stability of physical parameter region. However, in the cases around neutron star where
matter does not attain supersonic speed at all, sonic point does not come into the picture
of stability criterion. 

Now, when the sonic point exists in an accretion disk around compact object, $f_2(v_c,c_{sc})=0$ at $x=x_c$. Thus
at that location Mach number can be written as
\begin{eqnarray}
M_c^2=\frac{{\cal B}+\sqrt{{\cal B}^2-4\cal{A}\cal{C}}}{2\cal{A}},
\label{mac}
\end{eqnarray}
where
\begin{eqnarray}
\nonumber
{\cal A}&=&\alpha^2\gamma f(\Gamma_3-1)+\gamma(\Gamma_1+1)-\frac{4\alpha^2f(\Gamma_3-1)\gamma^2}{3\gamma-1},\\
\nonumber
{\cal B}&=&2\Gamma_1-\frac{4\alpha^2\gamma f(\Gamma_3-1)}{3\gamma-1}\left(1-\frac{I_{n+1}}{I_n}\right),\\
{\cal C}&=&\frac{2\alpha^2 f(\Gamma_3-1)}{3\gamma-1}\frac{I_{n+1}}{I_n}.
\end{eqnarray}
Subsequently, at $x=x_c$, $f_1(x_c,v_c,c_{sc})=0$. Thus, using (\ref{mac}) I can eliminate $v_c$ from $f_1(x_c,v_c,c_{sc})=0$
and get an algebraic equation for $c_{sc}$ and $x_c$, which I can solve to find out sound speed at sonic location.
To find out $\frac{dv}{dx}|_c$, I apply l'Hospital's rule to (\ref{dvdx}).
In order to understand the fluid properties in accretion disk, I have to solve (\ref{dvdx}) 
with an appropriate boundary condition. Also, integrating (\ref{azmom}), I get the angular momentum of the 
accreting matter as
\begin{eqnarray}
\lambda=\frac{\alpha x c_s^2}{v}\left(\frac{2}{3\gamma-1}+M^2\right)+\lambda_{in},
\end{eqnarray}
where $M$ is the Mach number of the flow and $\lambda_{in}$ is the angular momentum at the
inner edge of accretion disk.

Now from (\ref{ec}) and integrating (\ref{rmom}), I can write down the entropy and energy of the flow at sonic point as 
\begin{equation}
\dot{\cal M}_c=x_c^{3/2} F_c^{-1/2}(\gamma+1)^{q/2}\left(\frac{\frac{\lambda^2}{x_c^3}-F_c}{\frac{1}{F_c}
\frac{dF}{dx}|_c-\frac{3}{x_c}}\right)^{\frac{\gamma}{\gamma-1}}
\label{muc}
\end{equation}
and
\begin{equation}
E_c=\frac{2\gamma}{(\gamma-1)}\left(\frac{\frac{\lambda_c^2}{x_c^3}-F_c}{\frac{1}{F_c}\frac{dF_c}{dx}|_c
-\frac{3}{x_c}}\right)+V_c+\frac{\lambda_c^2}{2x_c^2},
\label{Ec}
\end{equation}
where $V_c=(\int F dx)|_c$ and $q=\frac{(\gamma+1)}{2(\gamma-1)}$. It can be mentioned that,
disk entropy and accretion rate are related by a simple relation as $\dot{\cal M}=(\gamma K)^n\dot{M}$.
While $\dot{M}$ is a conserved quantity for the particular accretion flow, $\dot{\cal M}$ is not,
as it contains $K^n$ which carries the entropy information that is not conserved in a dissipative system. 
As a boundary condition for the particular flow, I have to supply the sonic energy, $E_c$, and angular momentum, $\lambda_c$. 
Then from (\ref{Ec}), I can find out the sonic location $x_c$. Therefore, knowing $x_c$ one can
easily find out the fluid velocity and sound speed at the sonic point from $f_1=f_2=0$ with the 
help of (\ref{mac}) for a particular accretion flow. These have to be supplied as further boundary 
conditions of the flow.

\subsection{Sonic points in accretion disk}

Equation (\ref{Ec}) is a quartic equation of $x_c$
whose solution gives four roots, namely, sonic locations, of which either all are real or
two real and two complex or all are complex. In a particular flow,
if real roots at all exist, then at first it should be checked whether or not its location
is inside the black hole horizon ($x_+$) or stellar surface ($x_R$). 
The pseudo-potentials for the disk around
non-rotating compact object are finite for $x>2$. Thus the inner edge of the disk can be
extended upto $x=2$ for Schwarzschild geometry. However, the radius of neutron star
can be extended even upto $x\sim 6$ (at least theoretically).
If the root, $x_c>x_+,x_R$, sonic transition may be possible for that
flow. If there are two real physical roots, two possible
sonic transitions may exist. On the other hand if the roots are complex, there is no solution of
accretion flow for that particular parameter region having sonic point. Now from (\ref{dvdx}), 
it can be checked that numerator of $\frac{dv}{dx}|_c$ is the form of ${\cal Q}+
\sqrt{{\cal Q}^2-4{\cal P}{\cal R}}$ and 
I get the discriminant as $D={\cal Q}^2-4{\cal P}{\cal R}$ which determines the nature of
sonic points (also see Chakrabarti 1990). For any physical sonic transition $D>0$, where
for ${\cal C}>0$, ${\cal C}=0$ and ${\cal C}<0$, the sonic points are
`nodal', `straight line' and saddle (`X')-type respectively. When $D=0$, the sonic point is
`inflected nodal type' and for $D<0$, it is `spiral type' if ${\cal B}\ne 0$ and
`O(centre)-type' if ${\cal B}= 0$. Thus, there are two main kinds of sonic points
depending on $D$ is positive or negative, as $D=0$ (`inflected nodal type' sonic point)
is a special case of $D\ge 0$. All the sonic points for $D\ge 0$ are the distorted version
of `X'-type sonic point and for $D<0$ are basically the deviation of `O'-type sonic point.
If there are two saddle-type (`X'-type) sonic points in a particular
accretion disk, those form on either side of an `O'-type sonic point. If the saddle-type sonic
point forms at an inner radius than that of `O'-type sonic point, it is called the inner-sonic point
and if it forms at an outer radius, called outer-sonic point. The detailed discussions on
the definition and nature of various critical points
are given in Thompson \& Stewart (1986).

\subsection{Shock in accretion disk}

Now let me come to the issue of the formation of shock. Mukhopadhyay (2002b) showed that in accretion 
disk around neutron star, double shock is very natural in certain physical situations.
The shock wave may form in an accretion flow only when there are three critical/sonic
points exist (Chakrabarti 1989, 1996b). First, inflowing matter attains a supersonic speed at an outer sonic
point then it jumps from that outer sonic branch to inner sonic point branch by means
of a shock as the inner sonic branch has higher entropy, then passing through the inner
sonic point it attains supersonic speed again.
After that, if the compact object is a black hole, it falls into it. Otherwise, if it is a neutron
star, another shock forms close to the stellar surface and being subsonic matter, it falls
onto the surface of the neutron star.
There are three kinds of shock, namely, Rankine-Hugoniot shock, isentropic shock and isothermal
shock (Landau \& Lifshitz 1987; Chakrabarti 1990). 
In case of Rankine-Hugoniot shock, the energy is radiated away through the shock surface
and radiative cooling is inefficient. Here, $E_+=E_-$, $\dot{\cal M}_+>\dot{\cal M}_-$ and 
$T_+>T_-$. For isentropic shock, the energy is lost at the shock but the generation of 
entropy is comparable to the entropy radiated away keeping entropy conserved at the shock.
Here, $E_+<E_-$, $\dot{\cal M}_+=\dot{\cal M}_-$ and $T_+>T_-$. In the formation of isothermal
shock the radiative cooling is very efficient and some energy and entropy are lost at the
shock location keeping unchanged the temperature. Here, $E_+<E_-$, $\dot{\cal M}_+<\dot{\cal M}_-$
and $T_+=T_-$. Whatever be the nature of shock, the mass and momentum flux at the shock
location are always conserved. Thus, at shock location
\begin{equation}
\dot{M}_+=\dot{M}_-,\hskip1cm \bar{p}_++\bar{\rho}_+u_+^2=\bar{p}_-+\bar{\rho}_-u_-^2,
\label{sck1}
\end{equation}
where, '+' and '-' signs denote the quantities after and before the shock respectively and
$\bar{p}$ and $\bar{\rho}$ denote some kind of averaged pressure and density respectively
as the flow is considered in vertical equilibrium but with finite thickness.

However, in my discussion, whenever I mention the shock,  will mean Rankine-Hugoniot shock. 
If I write down the conditions to form a shock in an accretion disk (Chakrabarti 1989) for
rotating compact object, these are 
\begin{equation}
\frac{1}{2}M_+^2 c_{s+}^2+nc_{s+}^2=\frac{1}{2}M_-^2 c_{s-}^2+nc_{s-}^2,
\label{scke}
\end{equation}
\begin{equation}
\frac{c_{s+}^\nu}{\dot{\cal M}_+}\left(\frac{2\gamma}{3\gamma-1}+\gamma M_+^2\right)
=\frac{c_{s-}^\nu}{\dot{\cal M}_-}\left(\frac{2\gamma}{3\gamma-1}+\gamma M_-^2\right),
\label{sckmom}
\end{equation}
\begin{equation}
\dot{\cal M}_+>\dot{\cal M}_-,
\label{sckent}
\end{equation}
where
\begin{equation}
\dot{\cal M}=Mc_s^{2(n+1)}\frac{x_s^{3/2}}{\sqrt{F_s}}.
\label{sckent1}
\end{equation}
Here, $x_s$ indicates the shock location and $\nu=\frac{3\gamma-1}{\gamma-1}$.
Also from (\ref{sckmom}) and (\ref{sckent1}), I get the shock invariant quantity as
\begin{equation}
C=\frac{\left(\frac{2}{M_+}+(3\gamma-1)M_+\right)^2}{M_+^2(\gamma-1)+2}=
\frac{\left(\frac{2}{M_-}+(3\gamma-1)M_-\right)^2}{M_-^2(\gamma-1)+2},
\label{sckinv}
\end{equation}
which remains unchanged with respect to a non-rotating case. If only all the conditions, 
(\ref{scke})-(\ref{sckent}) and (\ref{sckinv}) are simultaneously satisfied by the matter, shock will
be generated in an accretion disk.

\section{Description of the disk parameter space}

I will show here, how does the rotation of compact object and viscous dissipation of the accreting
fluid affect the disk parameter region known for non-rotating compact object (according to Schwarzschild geometry). 
Therefore, I have to analyse globally the viscous accretion disk properties
around rotating black holes and neutron stars. 
Actually, I have to check, how do the angular momentum of compact object and viscosity of the 
disk fluid affect the sonic location, structure, as well as the stability of disk.
Also the magnetic field effect in the disk is considered to be negligible compared to the viscous effect. 
I will consider the rotation of black hole to be high as $J=\pm 0.9$, while '+' and '-' are 
for co-rotation and counter-rotation respectively. As it is known that 4U 1636-53 is one of the
fast rotating compact object other than black hole whose specific angular momentum is $J=0.2877$,
in my example for neutron star or compact object other than black hole, 
I consider the rotation to be $J=\pm 0.3$. I also will compare the results with non-rotating
cases. Along with inviscid case, I also will consider the viscous fluid with Shakura-Sunyaev
(1973) viscosity parameter ($\alpha$) chosen to be $0.4$ and $0.8$. When $\alpha=0.4$, the
heat lost is considered intermediate, as if half amount of the total viscous heat generated in the disk
is lost and thus cooling factor is chosen to be $f=0.5$. For $\alpha=0.8$, the heat lost is 
considered to be negligible as the residence time of the fluid in the disk is very small.
Thus the cooling factor is considered as $f=1$.

\subsection{Analysis for black hole}

Figures 1-3 are showing all the above mentioned effects in the disk around black hole. 
In Fig. 1a, I show the variation of disk entropy as a function of sonic location.
The intersections of all the curves by the horizontal line (which is a constant entropy line)
indicate the sonic points of the accretion disk for a particular entropy, viscosity of the disk 
and rotation of the black hole. It is clearly seen that, at a particular entropy, if the co-rotation of black hole
increases, marginally bound ($x_b$) and stable ($x_s$) orbits as well as sonic points shift to a more 
inner region and the possibility to have all four sonic points in the disk outside the horizon
increases. On the other hand, if the black hole counter-rotates, $x_b$ and $x_s$ move to greater
radii and all the sonic points come close to each other for the same entropy.
It is clearly seen from Figs. 1a-c that as the black is chosen to be rotating ($J=\pm 0.9$),
all the four sonic points appear outside to horizon, while for a non-rotating black hole
only three of them are appeared.
Now, if the viscosity of the accreting fluid increases keeping entropy unchanged, the possibility
to have all sonic points for a particular angular momentum of the black hole decreases.
In Fig. 1b, it is seen that as $\alpha$ increases from $0$ to $0.4$, for the counter-rotating 
black hole (here, $J=-0.9$), outer saddle- and centre-type sonic points have disappeared 
completely (as there is no intersection). Figure 1c shows, for a higher viscosity ($\alpha=0.8$),
even for co-rotating and no-rotating black hole, the outer saddle- and centre-type sonic 
points have disappeared.
Similar features are reflected from Fig. 2, where the sonic energy is plotted as a function of sonic
location. Here also, the intersections of the horizontal line (which indicates a constant energy line) 
by all the curves indicate the sonic points of the accretion disk for a particular energy, 
viscosity of the fluid and rotation
of the black hole. For both the Figs. 1 and 2, sonic points with negative slope of the curve indicate the
locations of 'saddle-type' sonic point and positive slopes indicate the 'centre-type' sonic point.
Thus the rotation of black hole and viscosity of the accreting fluid 
play the important roles to the formation and location of sonic points
which are related to the structure of accretion disk.


\begin{figure}
\epsscale{.80}
\plotone{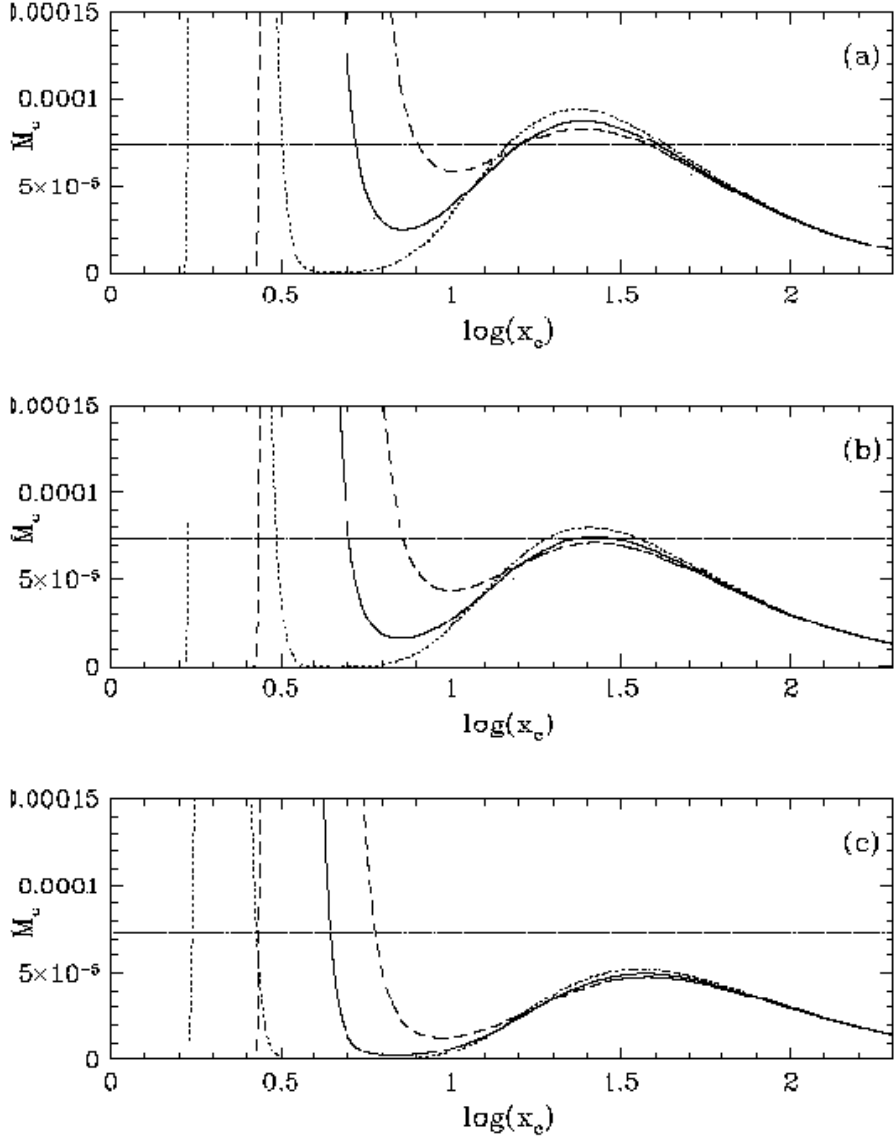}
\caption{
\label{fig1} Variation of sonic entropy as a function of sonic location in accretion
disk around black hole for various values of 
Kerr parameters, $J$, when (a) $\alpha=0$, $f=0$, (b) $\alpha=0.4$, $f=0.5$, (c) $\alpha=0.8$, $f=1$.
The solid curve indicates non-rotating case ($J=0$), while the dotted and dashed curves are
for co-rotating ($J=0.9$) and counter-rotating ($J=-0.9$) black hole respectively.
The horizontal line indicates the curve of constant entropy of $7.3\times 10^{-5}$.
$\lambda=3.3$, $\gamma=4/3$ for all the curves. 
}
\end{figure}

\begin{figure}
\epsscale{0.8}
\plotone{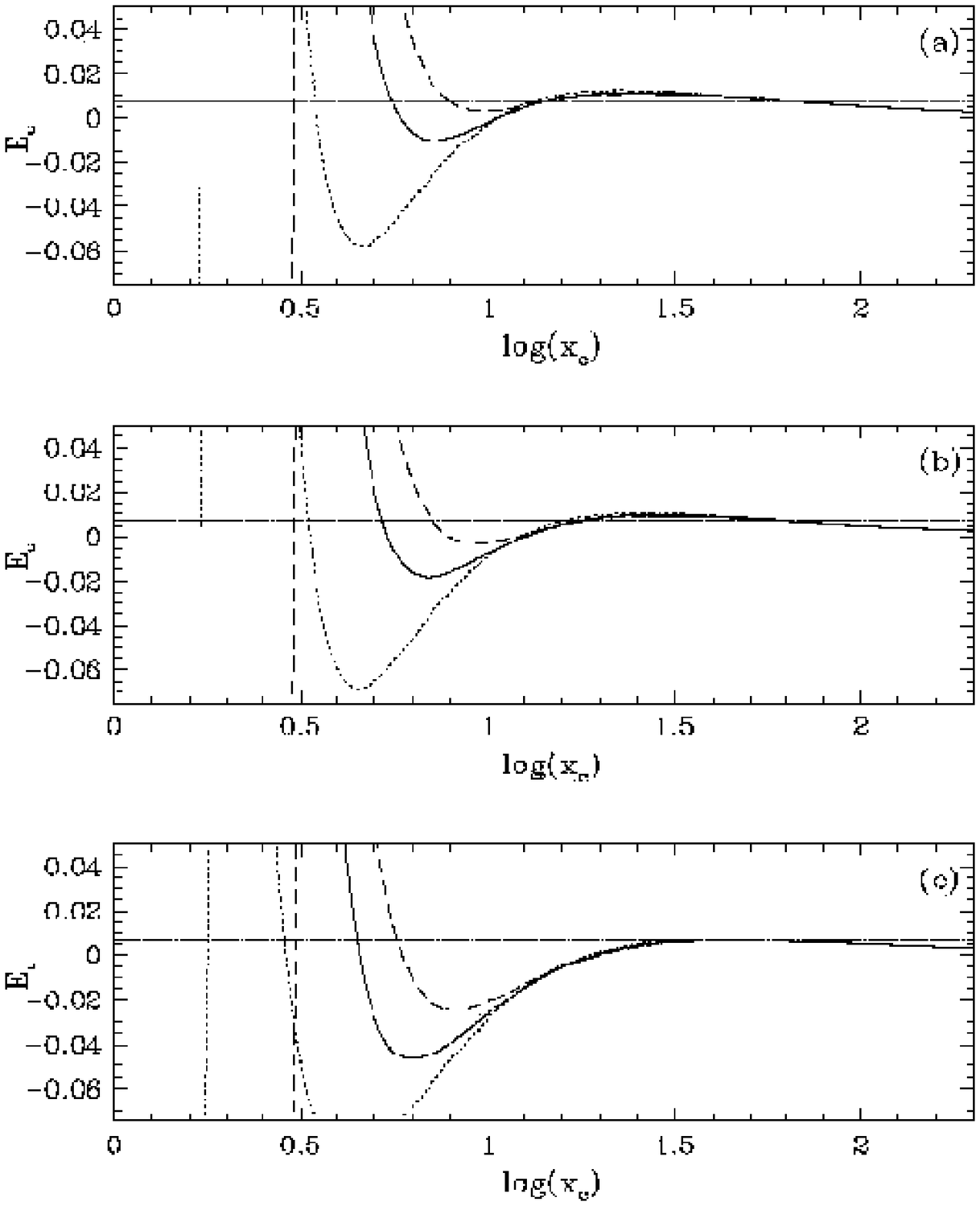}
\caption{
\label{fig2} Same as \ref{fig1}, except the sonic energy is plotted in place of sonic
entropy. The horizontal line indicates the curve of constant energy of $0.007$. 
}
\end{figure}


Figure 3 indicates the variation of energy with entropy at sonic locations, when $J$ is parameter
for different viscosity of the fluid. Here, 'I' and 'O' indicate the inner and outer sonic
point branch respectively. It is seen that, with the increase of co-rotation 
(for example, here $J=0.9$), inner sonic
point branch of the disk extends upto a lower entropy region keeping the outer sonic points unchanged
and the disk becomes unstable. I know that the shock in accretion disk can be formed, only if there is a
possibility of matter in the outer sonic point branch
of lower entropy to jump to the inner sonic point branch of higher entropy. As because, with
the increase of $J$, inner sonic point branch itself shifts to the lower entropy
region which is comparatively unstable, the shock as well as disk also become unstable.
Also with non-zero $J$, the inner and outer sonic point branches appear in such a manner that
there is no possibility of transition of matter from outer to inner sonic point branch
which increases entropy of the system. 
Clearly, for a highly rotating black hole, the possibility of formation of shock reduces as
well as the shock becomes unstable, as the inner sonic point branch itself is unstable.
The physical reason behind it is that, with the increase of value of $J$,
the angular momentum of the system increases, which helps the disk to maintain
a high azimuthal speed of matter upto very close to the black hole horizon, as a result the radial speed
of the matter can overcome the corresponding sound
speed at a very inner radius only. On the other hand, the inner edge of the accretion disk is comparatively
less stable, as the entropy decreases at lower radii. As the sonic points form at a more inner radii,
with the increase of $J$, disk tends to an unstable situation.

In case of the counter-rotation of black hole, as $J$ increases in magnitude,
the inner sonic point branch shifts towards a higher entropy region 
and the entire system shifts towards the outer side which is more stable.
Also I-branch merges or tends to merge to the O-branch and the possibility of shock decreases again.
The physical reason to have a more stable inner sonic
point branch for the counter-rotating case is the following.
As the counter rotation of central object increases, the angular momentum of the system
reduces, and the matter becomes more free to fall radially towards black hole at a 
larger radii and thus becomes supersonic
at a comparatively outer edge of the disk. Thus, the disk becomes more stable.
As in case of higher counter-rotation, the angular momentum of the system becomes very small, the corresponding
centrifugal pressure onto the matter becomes insignificant, as a result the possibility of shock
diminishes. Thus, if the angular momentum of black hole
increases or decreases significantly, the shock wave in accretion disk becomes unstable 
and the disk itself tends to an unstable situation. Also for high $J$, there is another 
centre-type sonic point branch appears outside the horizon which is seen at the right side 
of the I-branch (dotted and dashed curves) in Fig. 3. 

As the viscosity of flow increases, the possibility of intersection between inner and outer sonic
point branch decreases. With the increase of viscosity, more regions
containing the X-type sonic points become O-type (this feature is very prominent in Fig. 3c). 
The outer X-type sonic points recedes in further out
and inner one proceeds in further inwards. Thus the overall possibility of the formation of
shock reduces significantly.

\begin{figure}
\epsscale{0.9}
\plotone{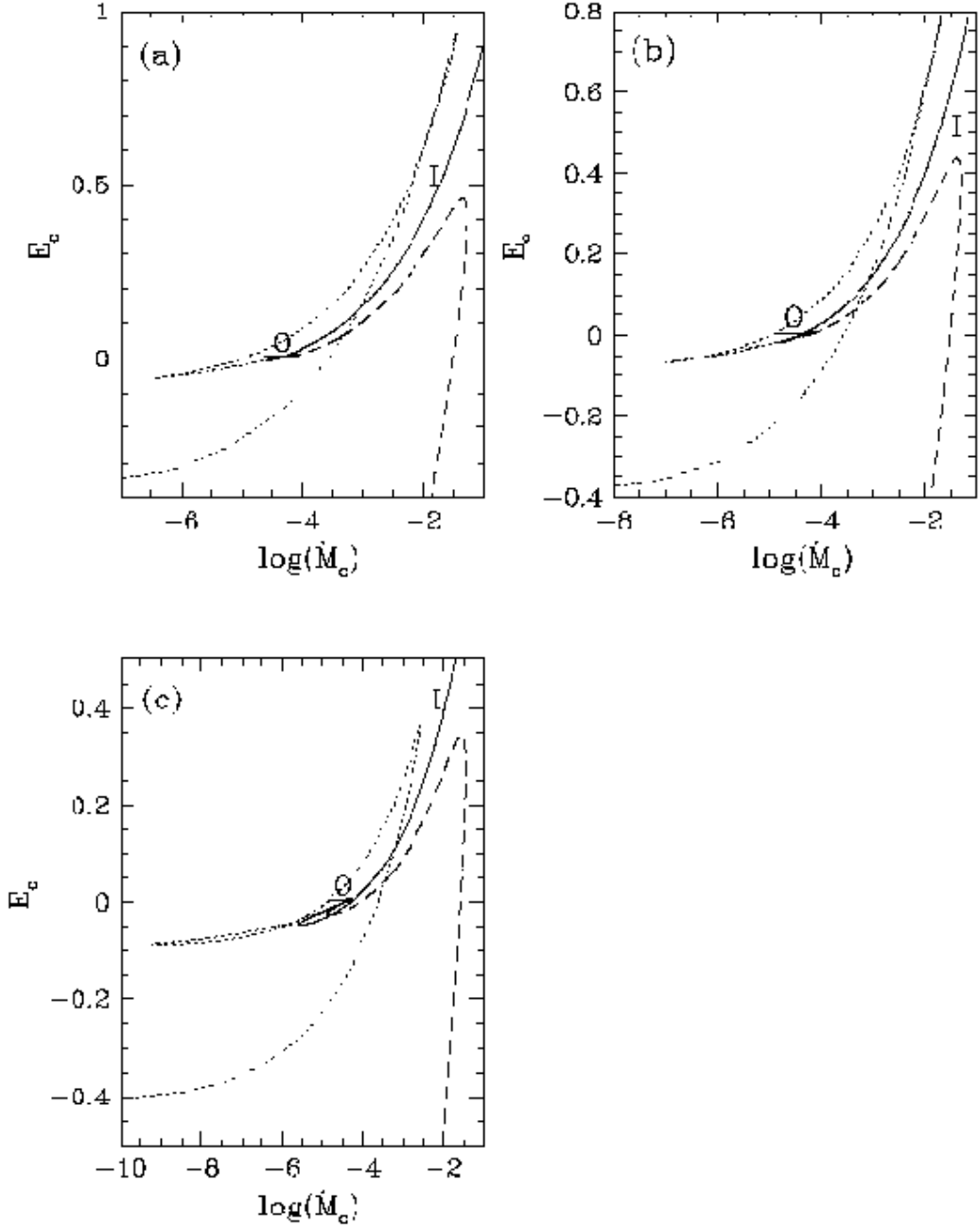}
\caption{
\label{fig3} Variation of sonic energy as a function of sonic entropy in accretion disk
around black hole for various values of
$J$ when (a) $\alpha=0,f=0$, (b) $\alpha=0.4,f=0.5$, (c) $\alpha=0.8,f=1$.
Solid curve indicates non-rotating case ($J=0$), while the dotted and dashed curves are
for $J=0.9$ and $-0.9$ respectively. 
O and I indicate the locus of outer and inner sonic point respectively for solid ($J=0$) curve.
The other parameters are $\lambda=3.3$, $\gamma=4/3$.
}
\end{figure}

\subsection{Analysis for neutron star}

In case of the accretion disk around rotating neutron star or other gravitating object which
has hard surface, close to the compact object the space-time will be different from that of
black hole. Thus the parameter space need not be same as of the disk around black hole 
at the inner edge. In Figs. 4 and 5, I show the variation of sonic entropy and
energy respectively as a function of sonic location. 
Figures 4a and 5a clearly show, while for non-rotating compact object three sonic points exist,
for a rotating case the number of sonic point may four. However, as the fourth sonic
point forms at a radius $x<2.6$ (see, Figs. 4 and 5), which is often inside the
stellar surface, practically even for rotation only three sonic points will form in
the accretion disk. In Figs. 4b,c and 5b,c, I show how the locations of sonic point
shift with the increase of viscosity of the accreting fluid. Figures 4c and 5c indicates
that for $\alpha=0.8$, practically the disk has only (physical, saddle-type) 
sonic point, namely inner one for any $J$. Other qualitative features are same as that
of black hole.

\begin{figure}
\epsscale{0.8}
\plotone{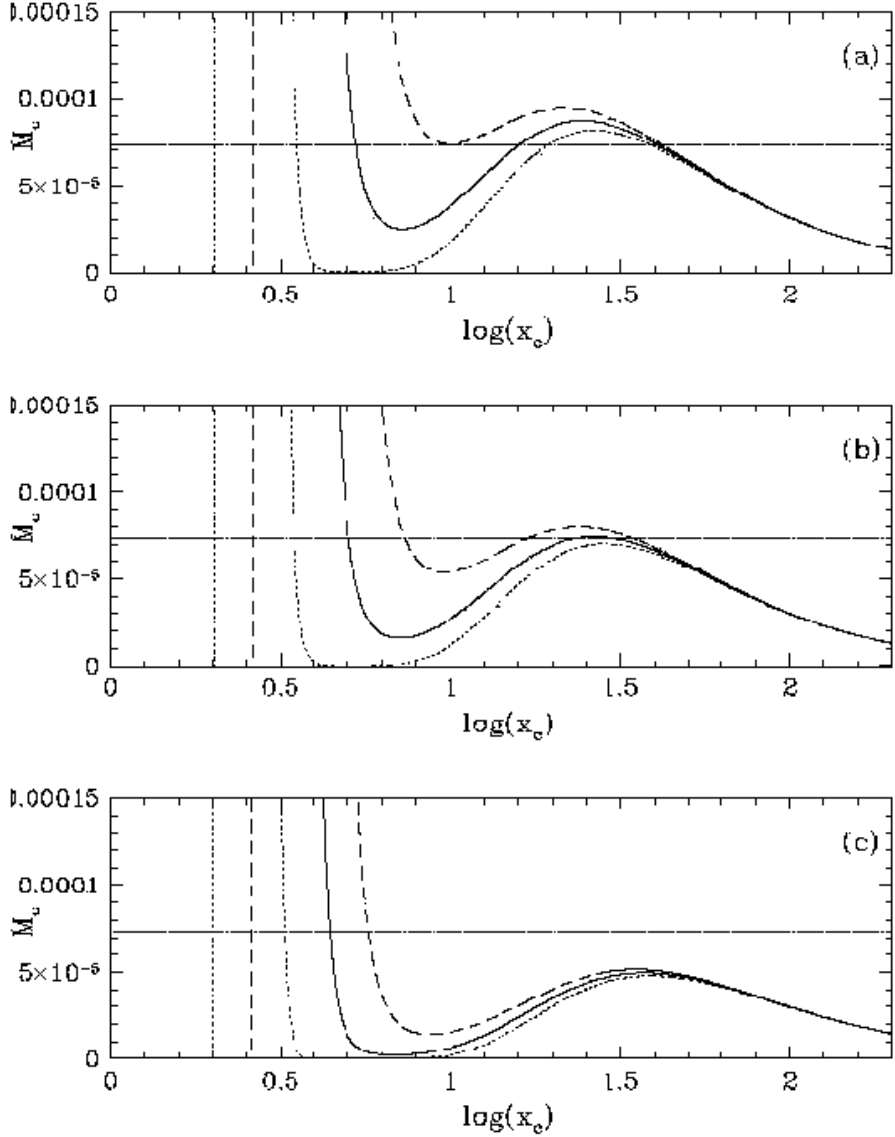}
\caption{
\label{fig4} Variation of sonic entropy as a function of sonic location in accretion
disk around neutron star for various values of 
angular momentum, $J$, when (a) $\alpha=0$, $f=0$, (b) $\alpha=0.4$, $f=0.5$, (c) $\alpha=0.8$, $f=1$.
The solid curve indicates non-rotating case ($J=0$), while the dotted and dashed curves are
for co-rotating ($J=0.3$) and counter-rotating ($J=-0.3$) compact object respectively.
The horizontal line indicates the curve of constant entropy of $7.3\times 10^{-5}$.
$\lambda=3.3$, $\gamma=4/3$, $M=2M_\odot$ for all the curves. 
}
\end{figure}

\begin{figure}
\epsscale{0.8}
\plotone{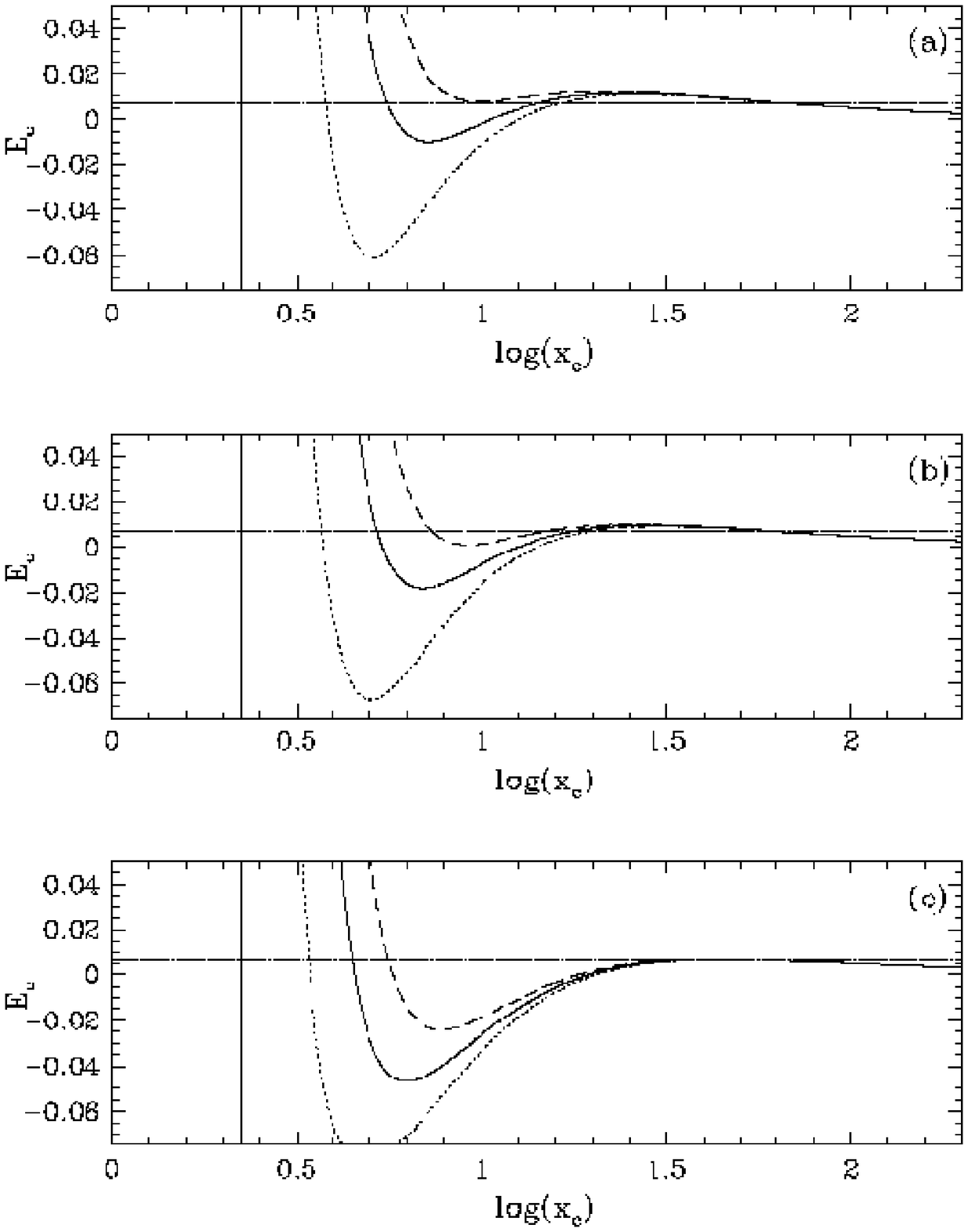}
\caption{
\label{fig5} Same as \ref{fig4}, except the sonic energy is plotted in place of sonic
entropy. The horizontal line indicates the curve of constant energy of $0.007$. 
}
\end{figure}

In Fig. 6, I show the variation of sonic energy as a function of sonic entropy of the disk,
when $J$ is parameter for different viscosity of the fluid. Different features and corresponding
physical reasons are similar to that of black hole.
For $J=0.3$, the inner sonic point branch disappears totally and only outer X-type and O-type
sonic point branches remain.
Clearly, for a (highly) rotating compact object, the possibility of formation of shock reduces as
well as the shock becomes unstable, as the inner sonic point branch itself disappears
or tends to disappear. In case of inviscid flow (Fig. 6a), for counter-rotating compact object,
inner and outer X-type (saddle type) sonic point branches merge each other. However, for viscous fluid,
although those are not merging but their clear distinction getting disappeared or seemed to
disappear and the branch of O-type sonic point tends to convert to a part of X-type branch. 
Overall, with the increase of viscosity all the sonic point branches tend to merge to a
single branch which results a single sonic point for a particular accretion flow.

\begin{figure}
\epsscale{0.8}
\plotone{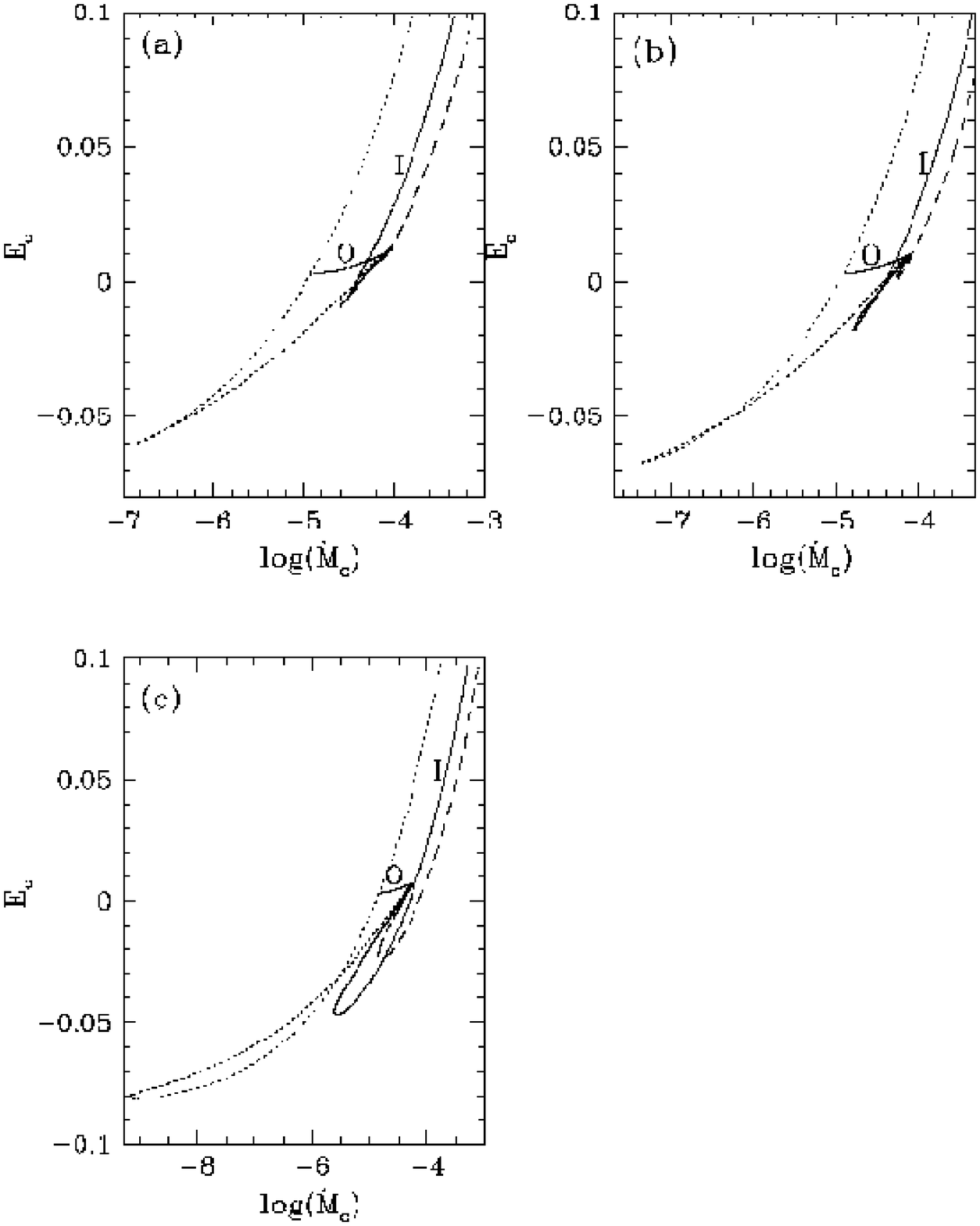}
\caption{
\label{fig6} Variation of sonic energy as a function of sonic entropy in accretion disk
around neutron star for various values of
$J$ when (a) $\alpha=0,f=0$, (b) $\alpha=0.4,f=0.5$, (c) $\alpha=0.8,f=1$.
Solid curve indicates non-rotating case ($J=0$), while the dotted and dashed curves are
for $J=0.3$ and $-0.3$ respectively. 
O and I indicate the locus of outer and inner sonic point respectively for solid ($J=0$) curve.
The other parameters are $\lambda=3.3$, $\gamma=4/3$, $M=2M_\odot$.
}
\end{figure}

\section{Fluid Properties of Accretion Disk }

I will now discuss the effect of rotation of compact object on the disk fluid. 
My intention will be to see the change 
of fluid properties known for the non-rotating case. I will consider the cases of 
inviscid as well as viscous fluid. I will separately consider the accretion disk around
black holes and neutron stars (or strange stars) and thus the corresponding pseudo-Newtonian 
potentials [Eqs. (\ref{M}) and (\ref{MG})] and will discuss both the cases in an unified scheme.

\subsection{Around Black Holes}

Here I will consider a few cases for the accretion disk around black holes. 
I will mainly consider the inviscid accreting fluid in the disk.

\begin{figure}
\epsscale{0.8}
\plotone{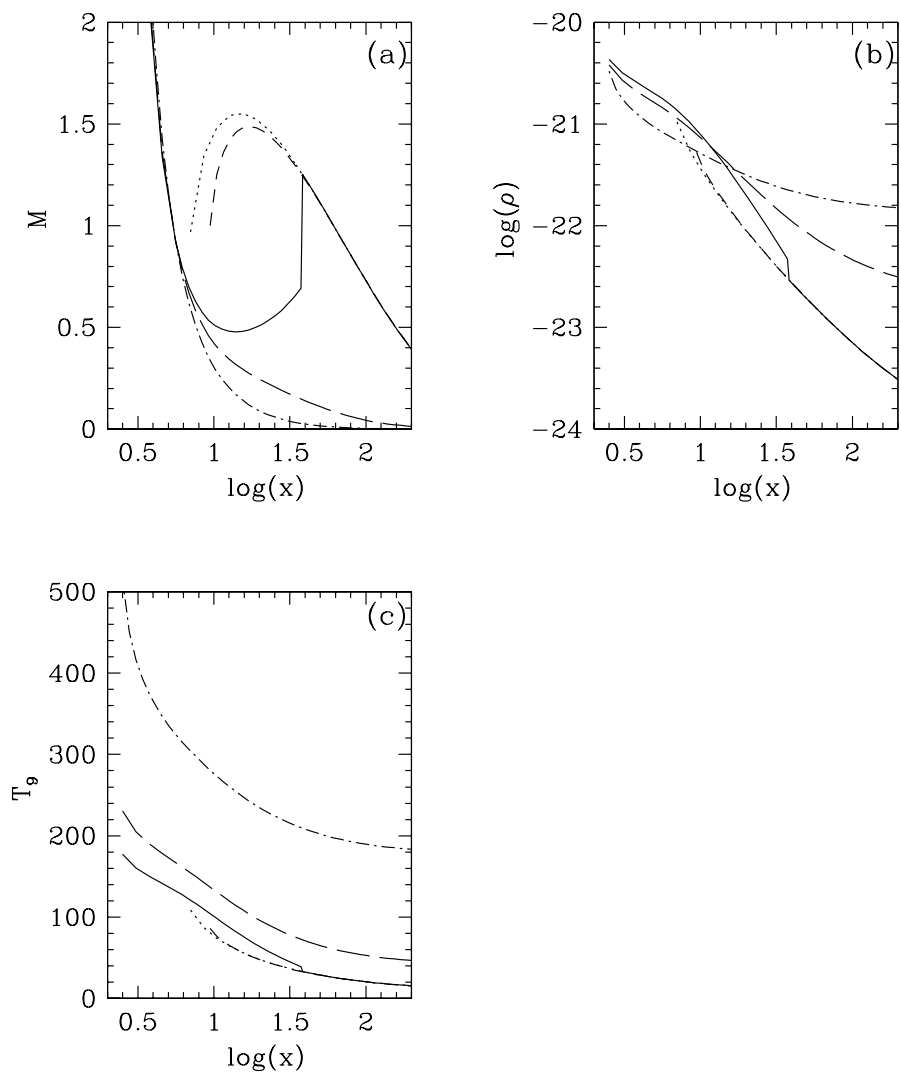}
\caption{
\label{fig7} Variation of (a) Mach no., (b) density and (c) temperature in unit of $10^9$ as a function of radial 
coordinate around black holes of $J=0$ (solid curve), $J=0.1$ (dotted curve), $J=0.5$ 
(dashed curve), $J=-0.1$ (long-dashed curve) and $J=-0.5$ (dot-dashed curve). Other parameters are,
$M=10M_\odot$, $\dot{M}=1$ Eddington rate, $\gamma=4/3$.
}
\end{figure}

Figure 7a shows the variation of Mach number for a particular set of physical parameter. If the black
hole is chosen to be non-rotating, shock forms in the disk. When the rotation 
of black hole is considered keeping other parameters unchanged, the 
solution changes and shock disappears. If the black hole co-rotates, the inner-edge of
disk becomes unstable and the matter does not find any physical path to fall steadily into the black hole. 
On the other hand, if the black hole counter-rotates, there is a single sonic point in the disk
and the matter attains a supersonic speed only close to the black hole and falls into it.
Thus, for the non-rotating and counter-rotating cases, there are smooth solutions of matter passing through
the inner sonic point, while for the co-rotating cases, it is not so.
The physical reasons behind these phenomena can be explained as follows. When the black hole co-rotates,
the angular momentum of system increases, the radial speed of matter may not be able to overcome the centrifugal barrier to fall into
the black hole for that parameter set. If the black hole counter-rotates, the angular momentum of 
system reduces, the centrifugal barrier smears out and matter falls steadily into the black hole. 
It can be mentioned here that, for the other choice of physical parameters 
(e.g. reducing/increasing the angular 
momentum of accreting matter for co/counter-rotation, changing the sonic locations etc.) 
the shock may appear again, even for the rotating
black holes (a few of such examples are depicted in Fig. 8).
Figure 7b shows the corresponding density profiles in unit of $\frac{c^6}{G^3M^2}$ for a particular accretion rate. 
As the matter slows down abruptly, the density of disk fluid jumps up at the shock location
for a non-rotating case. For the counter-rotating black holes,
the angular momentum of the system is less, thus, to overcome the centrifugal barrier, matter does not need 
to attain a high radial speed away from the black hole. In this situation, the radial 
matter speed is less and the
corresponding density of accreting fluid is high compared to a non-rotating case. I also show the
density profiles for co-rotating, unstable cases. In Fig. 7c, I compare the corresponding virial temperature
profiles in the disk in unit of $10^9$ ($T_9$). The variation of temperature is similar to that of density 
away from the black hole
but is opposite when close to the horizon (as the density varies inversely with the temperature 
at a small $x$ but not so at a large $x$, see Eqn. (\ref{ec})). As the higher counter-rotation
Mach number lowers down, the temperature of accretion disk becomes higher. 


\begin{figure}
\epsscale{0.8}
\plotone{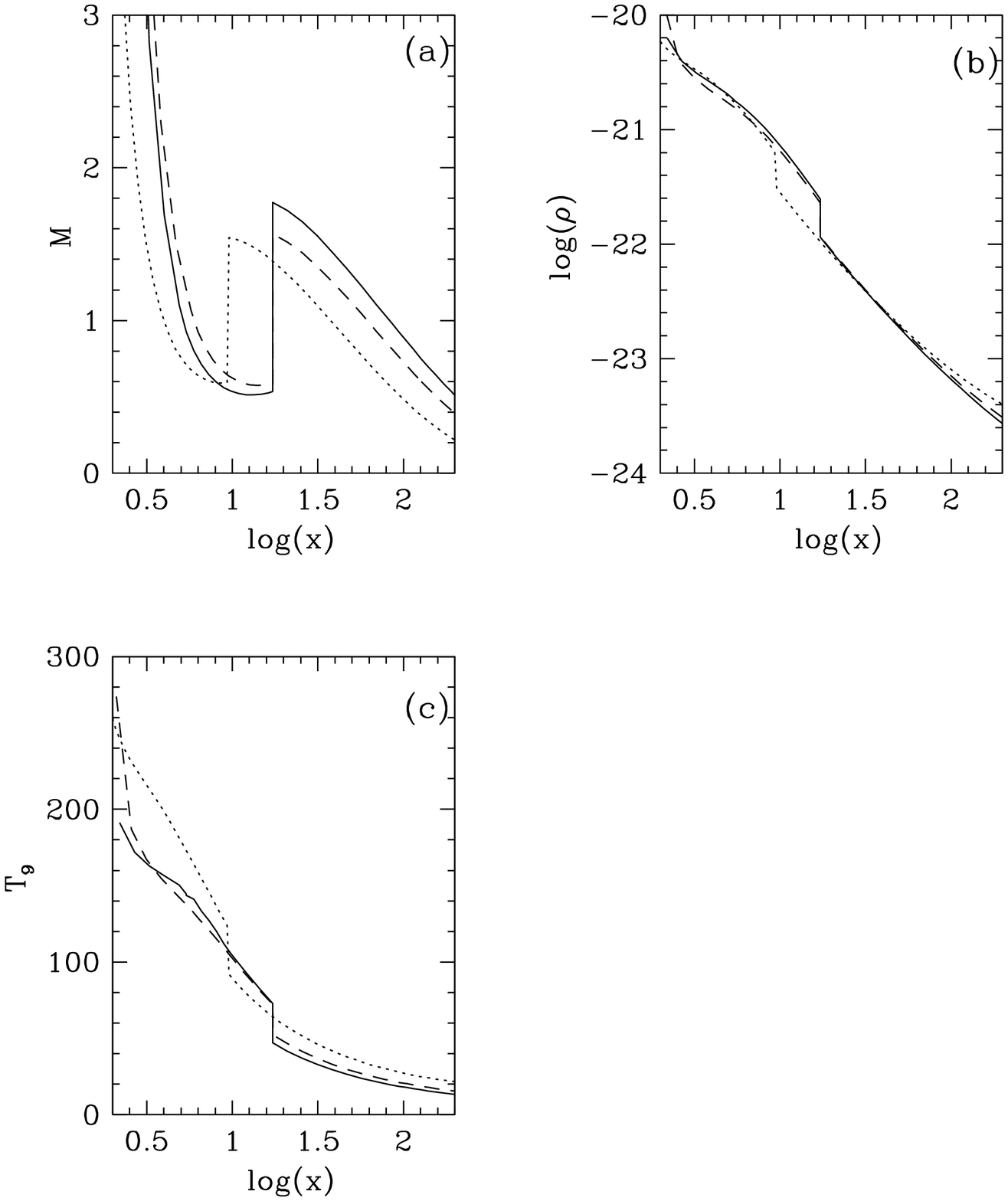}
\caption{
\label{fig8} Variation of (a) Mach no., (b) density and (c) virial temperature in unit of $10^9$
as a function of radial coordinate around black holes when the sets of physical parameters are 
(i) $J=0.1$, $x_o=94$, $x_i=5.4$,
$\lambda=3.3$ (solid curve); (ii) $J=0.5$, $x_o=43$, $x_i=4.2$, $\lambda=2.8$ (dotted curve) and 
(iii) $J=-0.1$, $x_o=70$, $x_i=6.324$, $\lambda=3.3$ (dashed curve). Other parameters are,
$M=10M_\odot$, $\dot{M}=1$ Eddington rate, $\gamma=4/3$.
}
\end{figure}

Figure 8 gives examples where the shock forms in an accretion disk around rotating black hole. 
The variation of Mach number for different Kerr parameters is shown in Fig. 8a. 
It reflects that, for a small rotation of black hole ($J=0.1$), 
matter adjusts the sonic locations in such a manner
that the outer ($x_o$) and inner ($x_i$) sonic locations shift in a more outer and more inner region 
respectively with respect to a non-rotating
case, and the shock forms in disk at $x=17.18$. For the counter-rotating case ($J=-0.1$), $x_o$ may remain
unchanged, but $x_i$ has to be shifted outside to form a shock at 
$x=17.25$ (that keeps the shock location (almost) unchanged).
As for the counter rotating cases, marginally stable ($x_s$) and marginally bound ($x_b$) 
orbits shift away from the black hole with respect to a co-rotating one, 
$x_i$ also shifts outside to stabilize the disk.
If the co-rotation of black hole is higher, say $J=0.5$, to form a stable shock, not only the inner sonic 
location has to be smaller, the disk angular momentum ($\lambda$) should also be reduced. 
High rotation of a black hole results in the location of horizon ($x_+$) as well as $x_b$ and $x_s$ 
to shift inwards, and the matter attains a supersonic speed at a point in more inside of the
inner edge of the disk to fall into the black hole.
Thus all the phenomena, like, sonic transitions, shock formations, etc., shift inside.
Moreover, for the stable solution, the angular momentum of the system can not be very high, otherwise
matter will be unable to overcome the centrifugal barrier. Thus, for the high rotation of a black hole,
like $J=0.5$, $\lambda$ has to be reduced to form a shock (at $x=9.51$) 
and stabilize the disk. However, for $J=-0.5$ and more negative,
there is no sub-Keplerian flow for which a shock may form. For the higher counter-rotation 
of a black hole, the angular momentum of system reduces. 
Since a higher centrifugal barrier is essential to form a shock, at the higher counter-rotation,
shock completely disappears in the accretion disk. 
Figures 8b and 8c show the corresponding variations of density and virial
temperature respectively. 

\begin{figure}
\epsscale{0.9}
\plotone{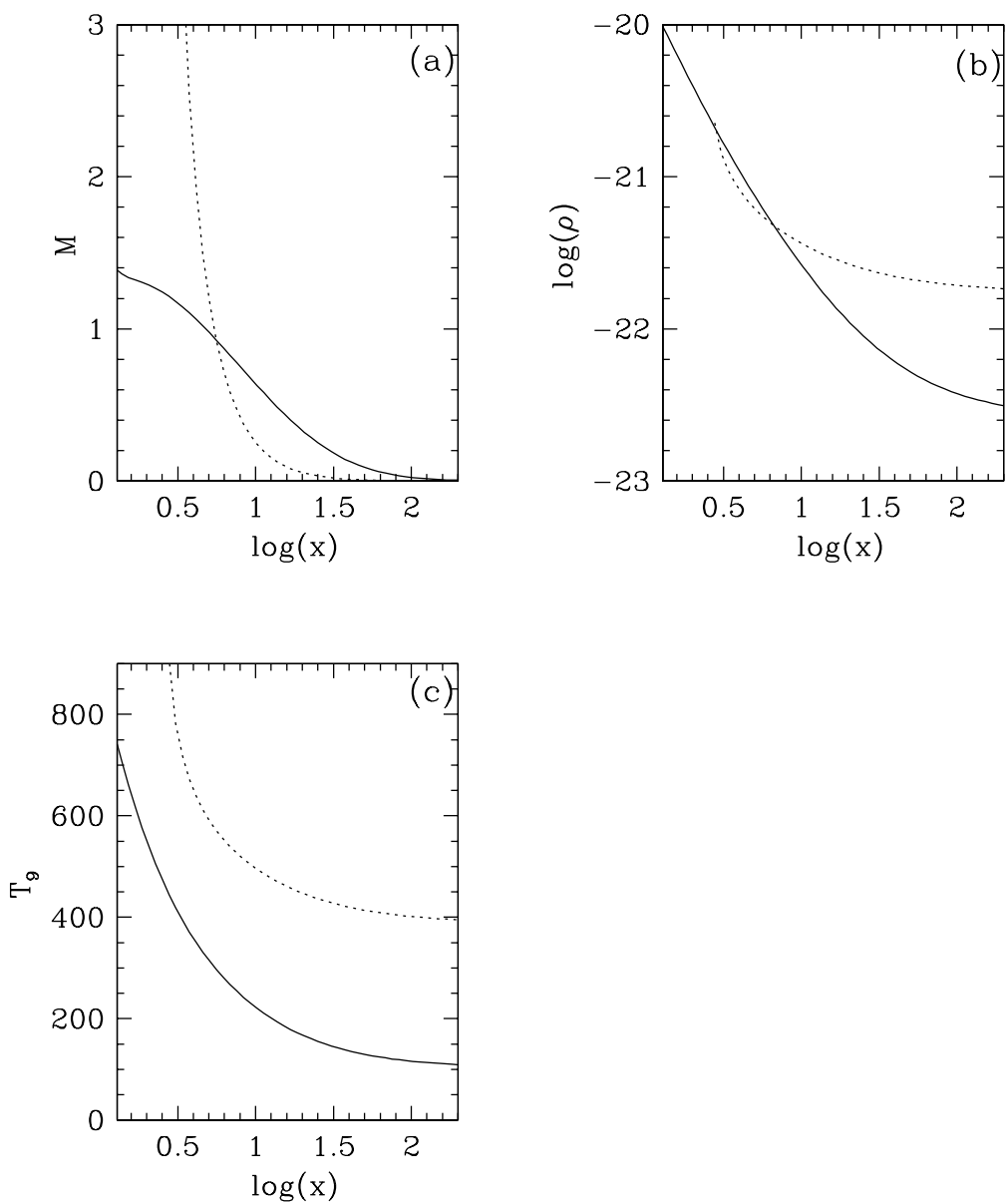}
\caption{
\label{fig9} Variation of (a) Mach no., (b) density and (c) virial temperature in unit of $10^9$ as a function of radial
coordinate around black holes for the sets of parameters (i) $J=0.998$, $x_i=5.6$, $\lambda=1.8$ (solid curve) and (ii) $J=-0.998$,
$x_i=5.6$, $\lambda=3.3$ (dotted curve). Other parameters are $M=10M_\odot$, $\dot{M}=1$ Eddington rate,
$\gamma=4/3$.}
\end{figure}

In Fig. 9, I show the fluid properties for $J=\pm 0.998$. At a very high rotation of black hole,
the possibility to form a shock in accretion disk totally disappears. For the co-rotating case,
the disk cannot be stable unless the angular momentum of accreting matter is very low 
(as I have already mentioned that the high angular momentum of the system makes the disk unstable). 
Therefore, there is no scope to form a shock, as to its formation, matter
needs a significant centrifugal barrier as well as a high angular momentum of the accreting fluid.
Here disk angular momentum chosen to be, $\lambda=1.8$ and the matter falls almost freely. For 
the counter-rotating case, the matter angular momentum need not be low to stabilize the disk. 
However, in the 
sub-Keplerian accretion disk, $\lambda$ cannot attain a high value to increase the
angular momentum as well as the centrifugal barrier of a system to form a shock. 
Thus, the shock disappears for a high counter-rotating case also. 
Figure 9a shows that $x_b$ and $x_s$ shift outwards for the retrograde orbit 
and the matter attains a supersonic speed at an outer radius than the case of a direct orbit in the disk. 
However, far away from black hole, the radial velocity is less (as it needs to 
overcome a little centrifugal barrier) than that of co-rotating case. 
Figure 9b shows the variation of corresponding density. Less velocity implies a high rate of
piling up of matter at a particular radius that produces a high density in the disk and thus the density 
behaves in an opposite manner to Mach number. Figure 9c shows the variation of virial
temperature in accretion disk. The temperature is low for the co-rotation and high for the counter-rotation. 
The physical reason behind it is as follows. For a co-rotation, the net speed as well as the kinetic 
energy of matter is less, producing 
 less temperature. On the other hand, for a counter-rotating case, the situation is opposite.
Although I show the virial temperature of the disk, in reality several cooling processes may take place 
that can reduce the disk temperature upto a few factors.

\subsection{Around Neutron Stars}

Now I will discuss the behaviour of viscous accreting fluid around rotating neutron star or
strange star considering the effect of magnetic field in the disk is negligible.
Similar to the case of disk around black holes, here it will be investigated that how the various 
fluid dynamical behaviour of an accretion disk are affected for various choice of $J$ and $\alpha$. 
For a certain physical parameter set, if the existence of sonic radii are possible in the accretion disk, 
the flow structures can be studied for that choice of sonic radii (or energy) of the accretion flow.

In Fig. 10a, I show the variation of Mach number for three different values of $J$ at a particular 
viscosity. As the viscosity is low, the rate of energy momentum transfer in matter is less. 
It is known that for an accretion around neutron star, in a certain cases, matter may be always subsonic
(Chakrabarti \& Sahu 1997; Mukhopadhyay 2002b; Mukhopadhyay \& Ghosh 2003) and at the stellar surface
matter speed reduces to zero. Owing to this fact, here such a situation is considered where
the speed of the accreting fluid is low, which accounts for the
high residence time of matter in the disk. The angular momentum of the disk is high at $x\sim 10$ 
(see Fig. 10d) and thus radial matter speed is low. Subsequently, though the disk angular momentum 
goes down, as the matter comes closer to the stellar surface, it starts to decelerate and the 
radial matter speed still remains low. This results to a higher the possibility of cooling, 
and thus cooling factor $f$ is chosen as $0.1$. As the energy momentum transfer rate in matter is low, 
the accretion rate is chosen to be intermediate. I have chosen a standard mass of compact star 
as $2M_\odot$. If the star is chosen to be non-rotating, angular momentum of the system is less and 
the centrifugal barrier is minimum. With the increase of $J$, this centrifugal barrier increases, 
and it becomes very high particularly for $J=0.5$. Mach number profiles clearly indicate that at the 
surface of compact star, accreting fluid stops. Also from Fig. 10b and 10c,
it comes out that sudden deceleration of the accreting fluid gives rise to a sudden enhancement 
of density and temperature close to stellar surface. Here, the temperature is thought to be cooled 
down by a factor of $1/30$ through the inverse-Compton effect in the radiation pressure dominated 
relativistic flow. Thus, I choose $\beta\sim 0.03$.
Nevertheless, the temperature is still very high. The density profiles are plotted in unit of 
$\frac{c^6}{G^3M^2}$. 

\begin{figure}
\epsscale{0.9}
\plotone{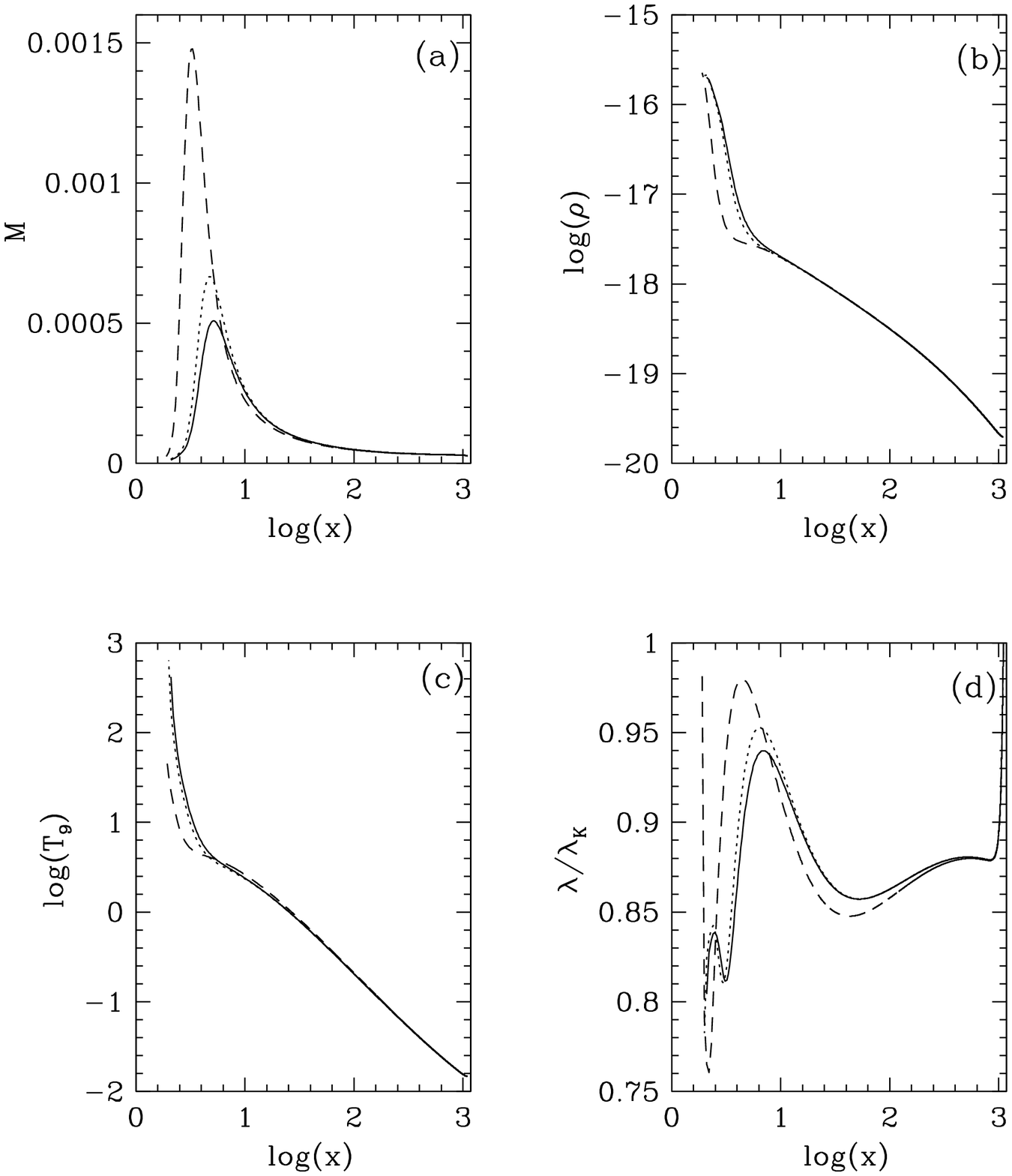}
\caption{
\label{fig10} Variation of (a) Mach number, (b) density, (c) temperature in unit of $10^9$, 
(d) ratio of sub-Keplerian
to Keplerian angular momentum of the accreting fluid around neutron stars as a function of radial coordinate. Solid,
dotted and dashed curves are for $J=0,0.1,0.5$ respectively. Other parameters are $\alpha=0.0001$, $f=0.1$,
$M=2M_\odot$, $\dot{M}=2$ Eddington rate, $\beta=0.03$.
}
\end{figure}

\begin{figure}
\epsscale{0.9}
\plotone{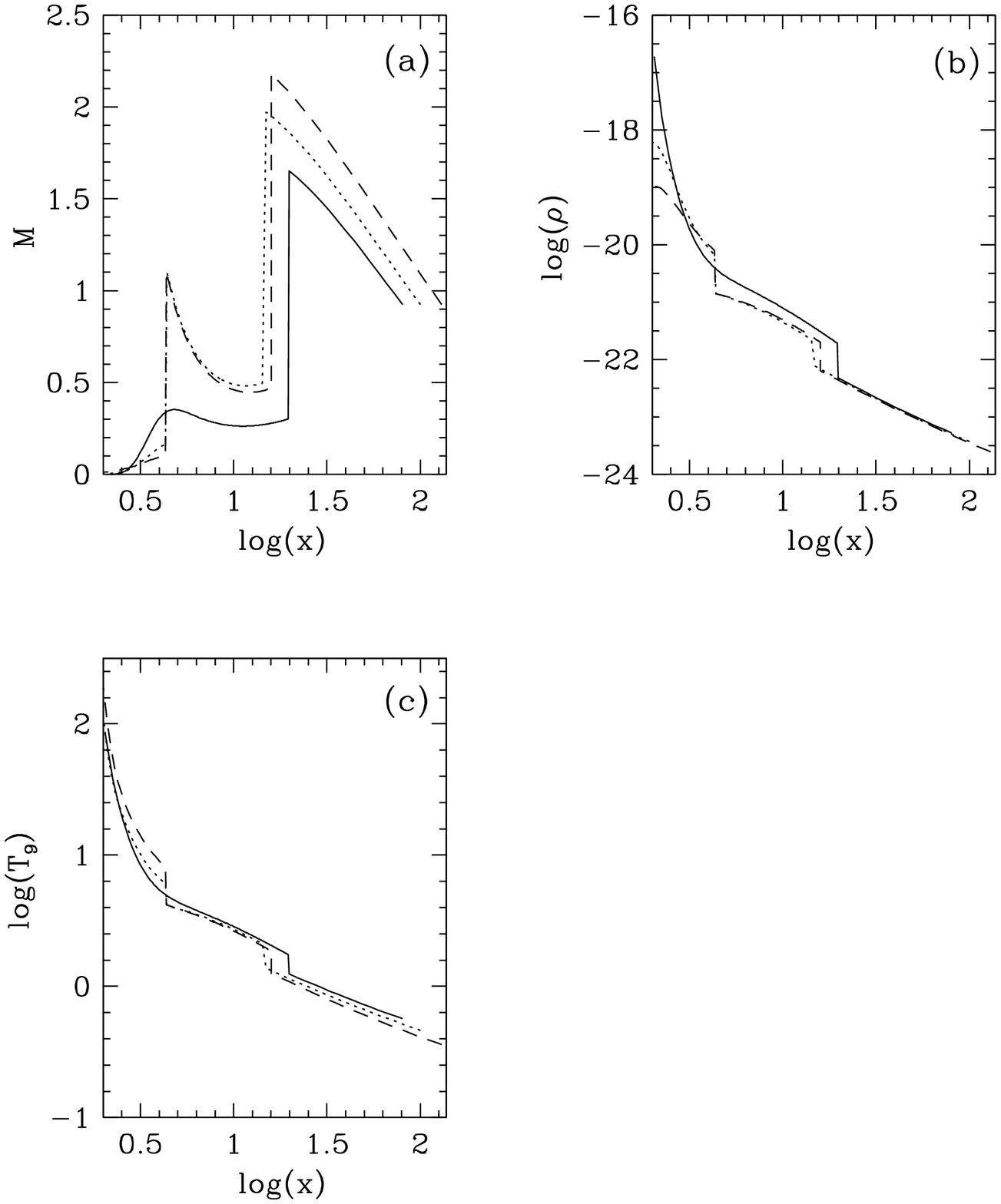}
\caption{
\label{fig11} Variation of (a) Mach number, (b) density, (c) temperature in unit of $10^9$
as a function of radial coordinate for the accretion disk around 4U 1636-53. Solid, dotted 
and dashed curves are respectively for
(i) $\alpha=0, f=0$, (ii) $\alpha=0.02, f=0.2$, (iii) $\alpha=0.05, f=0.5$.
Other parameters for 4U 1636-53 are $J=0.2877$, $\lambda_c=3$, $M=1.4M_\odot$, $\dot{M}=2$ 
Eddington rate, $\beta=0.03$.
}
\end{figure}

Figure 4d shows the ratio of sub-Keplerian to Keplerian angular momentum variation of the accreting
fluid as a function of radial coordinate. It is seen that, at the boundary between Keplerian and
sub-Keplerian disk, $\lambda_k/\lambda\rightarrow 1$, and from that radius ($x_K$) I start my calculation 
as my discussion is for sub-Keplerian accretion disk. From the $\lambda_k/\lambda$ profile, it is
again clear that for higher $J$, centrifugal barrier becomes stronger. 
Far away from compact star, the effect of centrifugal
barrier is weak, but as the matter approaches towards the compact object, 
it starts to dominate and at $x\sim 10$ it
becomes stronger. Then again it goes off. However, at very close to the stellar surface, 
again this centrifugal effect comes into the picture as the matter falls into the strong 
rotational field of compact object. This effect appears very prominently for high value of $J$, 
which is chosen to be $0.5$. Actually, this effect comes into the
picture, once the matter enters into the corresponding ergosphere of HT geometry. But the important 
point to be noted that if the outer radius of compact object is greater than the radius of 
ergosphere, this feature does not have scope to take place.

 In Fig. 11, I show the results of accretion phenomena around one of the fast rotating compact
 object, namely 4U 1636-53, whose angular momentum, $J=0.2877$.
 The examples are shown, where the sonic point(s) exist in the flow and shock forms in the accretion disk
 around 4U 1636-53 for different
 viscosity parameters. Figure 11a shows the variation of Mach numbers as a function of radial coordinate.
 It reflects that, for a small $\alpha$ ($\sim 0$) only shock is possible in the accretion 
 disk. As $\alpha$ increases, the energy momentum transfer rate increases, rate of infalling the matter 
 enhances and velocity of the accreting fluid becomes high at the inner edge of the disk. But, 
 because of the inner hard surface, matter has to stop at close to stellar surface and naturally 
 another shock forms at the inner edge of the accretion disk. The shock locations for $\alpha=0.02$ 
 and $0.05$, are at $x=14.27, 4.38$ and $x=15.37, 4.31$ respectively. If the viscosity is high, 
 the cooling factor is intermediate (see Fig. 5 caption).
 Also, the angular momentum of matter at the sonic point is chosen to be, $\lambda_c=3$, and for the 
 radiation dominated, inverse-Comptonised flow, $\beta$ is chosen to be $0.03$. Figures 11b and 11c 
 show the variations of density and temperature respectively.
  The regions, where velocity decelerates abruptly, temperature and density also jumps up in a 
  significant manner. Here also, the inverse-Comptonised temperature is very high as the cases of 
  Fig. 10.

\subsection{Comparison between Black Holes and Neutron Stars}

Here I will show a few examples, where the solution (mainly Mach number of accreting fluid)
changes with respect to the inner boundary conditions, namely, nature of compact object,
whether it is black hole or neutron star (or strange star). In this respect viscous 
accreting fluid is chosen and central black holes and neutron stars are chosen to be
non-rotating or in the limit of weak rotation.

\begin{figure}
\epsscale{0.85}
\plotone{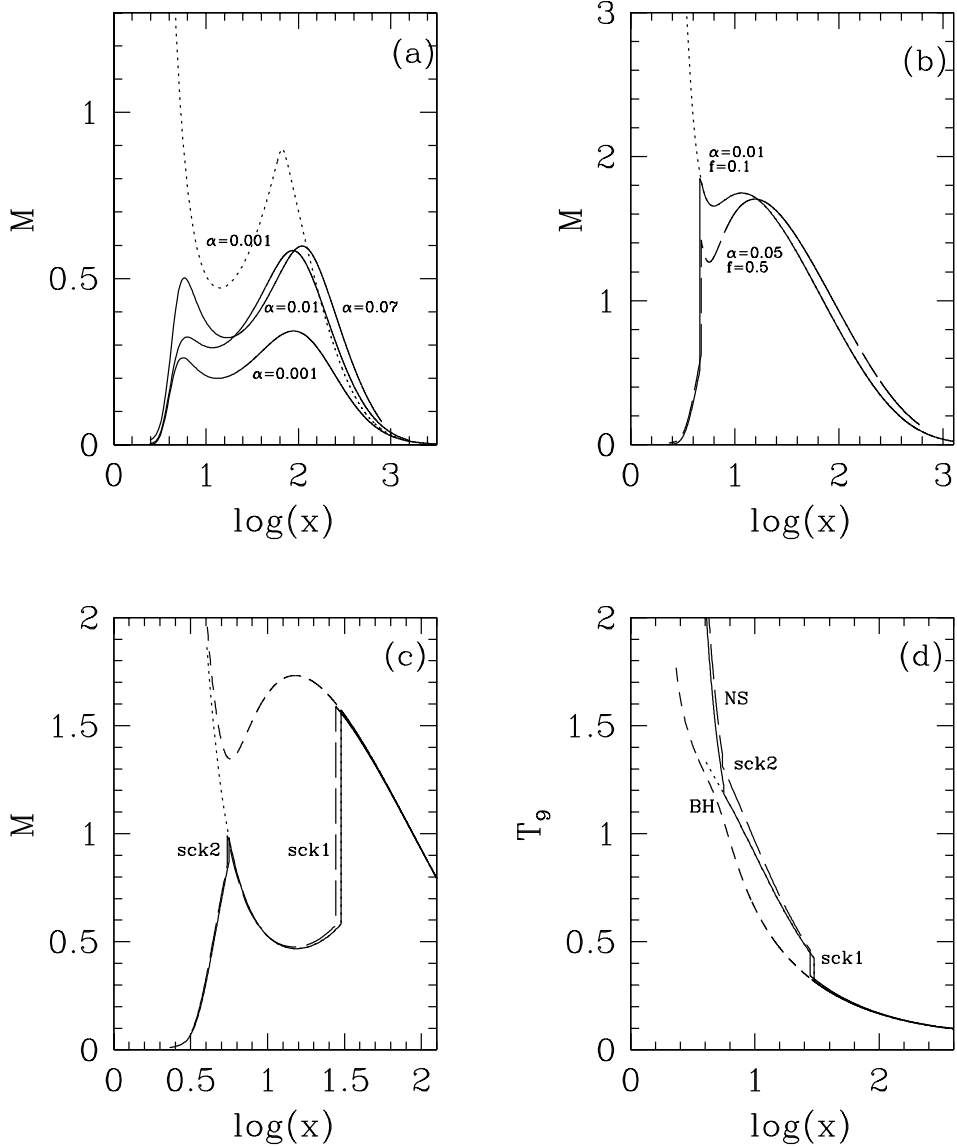}
\caption{
\label{fig12} Variation of Mach number, $M$, as a function of logarithmic radial distance when
(a) solid curves are for neutron star and dotted one is for accretion
around black hole; 
(b) solid and long dashed curves are for neutron star, dotted curve is 
for accretion around black hole; (c) solid ($\alpha=0.07,f=0.1$) and long-dashed ($\alpha=0.05,f=0.5$)
curves are for neutron star and dotted (stable case) and short-dashed (unstable case) curves 
($\alpha=0.07,f=0.1$) are drawn for accretion around
black hole. (d) Variation of temperature in unit of $10^9$ for the 'cases of (c)' 
as a function of logarithmic radial distance. By NS and BH, I mean the temperature distribution around
neutron star and black hole respectively. The shock locations are indicated as sck1 and sck2.
Others parameter for all the cases are, mass of the neutron star, $M_s=2M_\odot$, mass of the black hole,
$M_{BH}=10M_\odot$, $\gamma=4/3$, $\dot{M}=1$ Eddington rate, $\lambda_c=3.3$ when
$\alpha=0.001$, and $\lambda_c=3.2$ for others $\alpha$.
}
\end{figure}

Figure 12(a) shows the cases where the accreting fluid around neutron stars always stays 
in the subsonic branch, while around black hole (for $\alpha=0.001$) it must be supersonic
for the same parameter set. As $\alpha$ decreases, outer edge of the sub-Keplerian disk, $x_K$, increases,
as well as residence time of the matter in the disk increases.
As $\alpha$  decreases, energy momentum transfer rate also decreases, as well as the corresponding 
Mach number and the centrifugal force decrease. Thus with the increase of $\alpha$, incoming
matter attains more speed as well as more centrifugal force in the outward direction.

In Fig. 12(b), situations are considered in such a manner that close to the surface
of the neutron star shock forms, while for black hole no such shock is possible to form.
This case is slightly unstable as matter attains 
a supersonic speed at far away from the compact object and continues to fall supersonically from
$x\sim 100$ to just before the shock in the inner region of the disk. Usually in
stable cases, shock forms at a much greater radius as is shown in Fig. 12(c).
Here, no shock forms at the outer radius and matter
smoothly falls supersonically. If the neutron star is there at
the centre, in the very inner region, matter satisfies the Rankine-Hugoniot shock conditions and jumps
discontinuously from the supersonic to the subsonic branch
and falls on to the surface of neutron star, while for the case of central black hole matter continues
its supersonic motion and falls into the black hole. 

In Fig. 12(c), the parameters are considered in such a manner that the shock forms twice in
the flow around the neutron star, while for black hole a single shock is possible for a particular
physical parameter set. At first the matter satisfies the Rankine-Hugoniot shock conditions
at an outer radius irrespective of the nature of central compact object.
The formation of shock around black hole and neutron star at this location are similar. 
Subsequently, passing through the inner sonic point matter around black hole and 
neutron star again attains a supersonic speed. Now the formation of another shock depends on the
nature of compact object. If it is a black hole
there is no question of the second shock as the speed of the
matter must be supersonic close to the horizon. On the other hand, if there
is a neutron star, without having another shock matter can not
reach the surface of the star in this present situation.
This second shock occurs exactly in a similar way
as the shock in Fig. 12(b), if the corresponding Rankine-Hugoniot shock conditions are
satisfied. In Fig. 12(c) the shock locations are denoted by sck1 and sck2. 
I also show the Mach number variation
for the noshock unstable case by short-dashed curve, if the matter with $\alpha=0.07$ had fallen towards
a black hole of mass $10M_\odot$. Thus, the formation of
two shocks depends upon the nature of the
compact object and the possibility of satisfaction of shock conditions.
For certain choices of the flow parameter values, even for a neutron star
the Rankine-Hugoniot shock conditions
may not satisfy, though the matter attains a supersonic speed. For these
cases, the matter will not reach the neutron star.
It should be reminded that the inner shock location must be outside
the radius of the neutron star. Here, as the radius of the shock location
($\sim 5.4-5.6$) is greater than the usual radius of a neutron star of mass
$2M_\odot$ (Cook et al. 1994; Dey et al. 1998), inner shock forms safely.

Figure 12(d) shows the temperature variation in an accretion disk around the neutron
star and black hole for the cases shown in Fig. 12(c). Solid and long-dashed curves indicate the
temperature profiles around a neutron star (indicated
as NS in the figure) when viscosities of the infalling matter are $0.07$ and
$0.05$ respectively. At the shock locations (as indicated by sck1 and sck2 in figure) temperature rises
discontinuously. By the dotted and short-dashed curves I show the temperature
variation of the inflowing matter of $\alpha=0.07$ when it falls
towards a black hole of mass $10M_\odot$ (as indicated BH in the figure).
Dotted curve is drawn for stable shock case and short-dashed curve is
for unstable no-shock case (if the same matter had fallen
without forming a shock). Although the virial temperature of accretion
disk may be very high as of the order of $10^{11}$K, following Mukhopadhyay \& Chakrabarti (2000)
I have taken into account the various cooling processes so that the temperature
reduces to of the order of $10^9$K. As I consider the entire flow is
relativistic, i.e., $\beta$ is very low (radiation pressure dominated flow),
the soft photon in the disk is very
profuse. Thus, the temperature may be reduced by the inverse-Compton
effect. As there is the possibility of the formation of two shocks in the disk
around a neutron star, temperature is still high enough. It is very clear from the above discussions
that for the accretion around a neutron star, temperature and density are
higher compared to that around a black hole.

\section{Summary and Discussion  }

I have studied the viscous accretion phenomena around rotating compact objects. 
The corresponding accretion disks are modeled by means of Mukhopadhyay and MG 
pseudo-Newtonian potentials in order to describe the relativistic properties 
around black holes and neutron stars (strange stars).
As these potentials can describe all the essential relativistic properties 
like, radius of marginally bound and stable orbit, specific mechanical energy etc., of accretion disk
within $10\%$ error, I call the study as {\it pseudo-general-relativistic}, where, 
along with these potentials, I have described the accretion disk 
using non-relativistic equations. I have prescribed the set of basic equations
for vertically averaged, geometrically thin, viscous accretion disk and analysed the viscous parameter
space globally. Subsequently, I have discussed the properties of viscous fluid and the effect
of rotation of compact object on the fluid behaviour. 
Examples are shown for the accretion disk around black holes as well as an observed 
LMXB candidate 4U 1636-53, which is one of the fast rotating compact object with angular frequency 
582Hz (equivalent $J=0.2877$). I have treated the accretion disk around black holes and neutron
stars in a same approach and compared their properties. Just because of difference of inner boundary 
conditions, the accretion around a black hole and a neutron star differ each other 
at the inner edge of the disk.

As no cosmic object is static, one of my aim was to study the structure and 
stability of the accretion disk around rotating compact object and how does
the rotation play an important role in this respect, particularly
for the discussion of an inner edge of the disk. I have found that, when the rotation to a compact object
is incorporated, the locations of sonic point (if any) get shifted to the unstable region 
and the shock (if any) gets affected in the disk, and thus the disk structure gets influenced. 
The disk properties and phenomena not only get modified and/or shifted in location, 
sometimes disappear completely. Prasanna \& Mukhopadhyay (2003), worked on the stability 
analysis of accretion disk around rotating compact objects in a perturbative approach. They incorporated the 
rotational effect of central object in a disk indirectly by the inclusion of Coriolis acceleration term. 
Here, this rotational effect is brought directly from the metric itself.

From the global analysis of sonic points, it has been shown that for a higher co-rotation of compact
object, the disk becomes unstable for a particular angular momentum of the accreting matter. As the formation
of shock needs a stable inner sonic point in the accretion disk, for the higher value of $J$, shock is unstable
as the inner region of disk is unstable. 
In that regime, shock may disappear by any disturbance
created on the accreting matter and thus the matter may not get a steady inner sonic point.
On the other hand, for counter-rotating cases, the angular momentum of system reduces and the matter falls more 
strongly to a compact object. As there is no significant centrifugal barrier to slow
down the matter in disk, the possibility of shock reduces again. 
For a very high counter-rotation,
shock disappears completely. Also the branch of inner sonic point
merges or tends to merge to that of outer sonic point. Thus one can conclude that
the parameter region where the shock is expected to form (Chakrabarti 1989, 1990, 1996a) for non-rotating
compact objects, gets affected for the rotating ones. For a positive $J$, shock might have been formed
for a lesser value of the angular momentum of accreting matter. 
On the other hand, for a negative $J$,
to form a shock, the angular momentum of accreting matter should have a larger value than that
of non-rotating cases. As I am studying the sub-Keplerian accretion flow, the angular
momentum of matter cannot be unlimitedly high
(e.g., for a non-rotating black hole, $\lambda\lsim 3.6742$ at last stable orbit).
Thus, to form a shock, matter cannot attain the angular momentum beyond a certain limit. 
Also for a highly co-rotating compact object, to stabilize the disk, 
$\lambda$ has to be reduced by some physical process,
otherwise the radial speed could not overcome the centrifugal 
pressure to form a steady inner edge of the disk.
However, at an intermediate co-rotation, 
the parameter region of accretion disk enlarges to have three sonic points
and there is a possibility of a stable shock. On the other hand, for a counter-rotation, the valid
parameter region with three sonic points at the disk shortens and thus the possibility of shock reduces.
Similarly, for the higher viscosity, both the inner and outer
sonic point branch merge each other and the possibility of shock reduces again. 
Actually, with the increase of viscosity, the region containing the X-type sonic points tends to 
become O-type. The outer X-type sonic points recede in more outwards and inner ones proceed in inwards more. 

In case of compact object other than black hole, as the incoming matter slows down abruptly 
at the surface of compact object as well as at two
shock locations (if two shocks form), the overall density becomes higher
for an accretion disk around it compared to that for a black hole. Similarly, the temperature in the
disk is lower for a black hole. Thus, one can conclude that the accretion disk around a neutron star
is very favourable for nucleosynthesis. It was known that the disk around a black hole is enough hot 
for nucleosynthesis (Mukhopadhyay \& Chakrabarti 2000), which is different from that in star.
As the density and temperature may be much higher around a neutron star, 
more efficient nucleosynthesis is expected particularly at an inner region of the disk. In 
early, it was investigated that the high 
temperature of accretion disk around a black hole is very favourable for the photo-dissociations 
and the proton capture reactions (Mukhopadhyay \& Chakrabarti 2000, 2001). 
As the accretion disk around a neutron (or strange) star is hotter
than that around a black hole, even $^4\!He$, which has a high binding energy,
may dissociate into deuterium and then
into proton and neutron. If I consider the accreting matter to come
from the nearby 'Sun like' companion star, the initial abundance of
$^4\!He$ in the accreting matter is supposed to be about $25\%$. Therefore, by the dissociation of
this $^4\!He$, neutron may produce in a large scale, which could give rise to the
neutron rich elements. Guessoum \& Kazanas (1999) showed that the profuse
neutron may be produced in the accretion disk and through the spallation
reactions {\it lithium} may be produced in the atmosphere of the star.
When the neutron comes out from the accretion disk by the formation of
an outflow, in that comparatively cold environment,
$^7\!Li$ may be produced which can be detected on the stellar surface.
In early, it was shown that the metalicity of the galaxy may be influenced
when outflows form in the hot accretion disk around black holes (Mukhopadhyay \& Chakrabarti 2000).
In case of the lighter galaxy, the average abundance of the isotopes of
$Ca$, $Cr$ may significantly change.
Also the abundance of lighter elements, like the isotopes of $C$, $O$,
$Ne$, $Si$ etc. may increase significantly. As the
temperature of the accretion disk around neutron star is higher, the expected
change of abundance of these elements and the corresponding influence on
the metalicity of the galaxy is expected to be high.

It is known that, out of an observed pair of kilohertz QPO frequencies for a particular
candidate, one is the oscillation due to its Keplerian motion (Osherovich \& Titarchuk 1999). For 4U 1636-53,
lower frequency is 950Hz (Wijnands et al. 1997) which may be the Keplerian one. 
According to MM pseudo-potential (which presumably can describe the temporal effect in the disk), 
if I calculate this Keplerian frequency for 4U 1636-53, 
it comes out to be 927Hz for $M=1.4M_\odot$ and Keplerian radius, $x_K=16.5$, which is very close to
the observed one. Thus, I can give a theoretical prediction of QPO, at least for
one, out of a pair. Similarly, for other candidates, one can calculate Keplerian frequency
according to the pseudo-potential and compare with observation. The theoretical study of QPO
based on this scheme has been pursued by Mukhopadhyay et al. (2003).


\begin{acknowledgements}

\end{acknowledgements}

{}

\end{document}